\documentclass[a4paper,fleqn,usenatbib,useAMS]{mnras}

\usepackage{graphicx}
\usepackage{amsmath}	
\usepackage{amssymb}	
\usepackage{multicol}        
\usepackage{bm}		
\usepackage{pdflscape}	
\usepackage[encapsulated]{CJK}
\usepackage{ucs}
\usepackage[utf8x]{inputenc}
\usepackage{xcolor}

\usepackage[T1]{fontenc}
\usepackage{ae,aecompl}

\usepackage{txfonts}

\title[Dust in M\,1-67]{Dust in the Wolf-Rayet nebula M\,1-67}
\author[P.\,Jim\'{e}nez-Hern\'{a}ndez, S.J.\,Arthur and J.A.\,Toal\'{a}]{P.\,Jim\'{e}nez-Hern\'{a}ndez\thanks{E-mail:\,p.jimenez@irya.unam.mx}, S.J.\,Arthur and J.A.\,Toal\'{a}\\
Instituto de Radioastronom\'{i}a y Astrof\'{i}sica (IRyA), UNAM Campus Morelia, Apartado postal 3-72, 58090 Morelia, Michoacan, Mexico
}

\pubyear{2018}

\begin{document}
\label{firstpage}
\pagerange{\pageref{firstpage}--\pageref{lastpage}}
\maketitle

\begin{abstract}

  The Wolf-Rayet nebula M\,1-67 around WR\,124 is located above the
  Galactic plane in a region mostly empty of interstellar
  medium, which makes it the perfect target to study the mass-loss
  episodes associated with the late stages of massive star evolution.
  Archive photometric observations from \textit{WISE},
  \textit{Spitzer} (MIPS) and \textit{Herschel} (PACS and SPIRE) are
  used to construct the spectral energy distribution (SED) of the
  nebula in the wavelength range of 12--500~$\mu$m.  The infrared
  (photometric and spectroscopic) data and nebular 
  optical data from the literature are
  modeled simultaneously using the spectral synthesis code
  \textsc{Cloudy}, where the free parameters are the gas density
  distribution and the dust grain size distribution. The infrared SED
  can be reproduced by dust grains with two size distributions: a MRN
  power-law distribution with grain sizes between 0.005 and
  0.05~$\mu$m and a population of 
  large grains with representative size
  0.9~$ \mu$m. The latter points towards an eruptive origin for the
  formation of M\,1-67. The model predicts a nebular ionized gas mass of
  $M_\mathrm{ion} = 9.2^{+1.6}_{-1.5}~\mathrm{M}_\odot$ and the estimated mass-loss
  rate during the dust-formation period is $\dot{M} \approx 6 \times
  10^{-4}~ \mathrm{M}_\odot$~yr$^{-1}$.  We discuss the implications
  of our results in the context of single and binary stellar evolution
  and propose that M\,1-67 represents the best candidate for a
  post-common envelope scenario in massive stars.
  
  \end{abstract}





\section{INTRODUCTION}
\label{sec:intro}

Wolf-Rayet (WR) stars represent the most advanced evolutionary stage
of very massive stars prior to supernova explosion \citep[$M_\mathrm{i} 
\gtrsim25$~M$_{\odot}$;][]{Langer1995,Ekstrom2012}. Mass loss is important at all stages of a massive star's life and the WR stage corresponds to the He-rich core once the outer envelope has been stripped away. 
In the single massive star scenario, O stars with initial masses $25~\mathrm{M}_\odot < M_\mathrm{i} 
< 60~\mathrm{M}_\odot$ evolve into a red supergiant (RSG) stage exhibiting 
slow (10--30~km~s$^{-1}$) and dense winds ($\dot{M} \approx10^{-4}$~M$_{\odot}$~yr$^{-1}$), 
whilst stars with even higher initial masses ($> 60 M_\odot$) are thought to 
pass through a luminous blue variable (LBV) stage \citep{Georgy2012}, 
in which they experience eruptive mass-loss episodes 
\citep[$\dot{M}\gtrsim10^{-3}$~M$_\odot$~yr$^{-1}$;][]{Weis2001}. 
Finally, after losing its hydrogen-rich envelope through the 
RSG or LBV phase, the star exposes its hot 
core becoming a WR star. However, the effect of binarity on stellar evolution cannot be ignored, since most O stars are born in close binary or multiple systems \citep{Mason2009}. During the lifetime of the system, there will be some interaction before the massive star ends its life as a supernova \citep{Sana2012}. In the binary scenario, Roche lobe overflow can lead to enhanced mass loss of the primary and mass transfer to the secondary, altering the lifetimes of both components in the different evolutionary stages \citep[e.g.,][]{VandenHeuvel1972,deMink2014ApJ}. One
of the components might have its hydrogen-rich 
envelope stripped completely, thereby creating a WR star
\citep[e.g.,][]{Wellstein2001}.

Regardless of the WR formation process, their powerful
line-driven winds ($v_{\infty} >1000$~km~s$^{-1}$,
$\dot{M}\approx10^{-5}~\mathrm{M}_{\odot}$~yr$^{-1}$; \citealp{Hamann2006}) and
high ionizing photon luminosity ($Q_\mathrm{H} >                                
5\times10^{48}$~s$^{-1}$) interact with the circumstellar environment
producing optical WR nebulae in many cases. 
Thus, the study of WR nebulae can provide information on the mass-loss history and previous evolution of the
central star.

WR nebulae have been characterized through optical images and
spectra. Narrow-band filter images have been used to classify their
[O\,{\sc iii}], H$\alpha$ and [N\,{\sc ii}] morphologies into three
broad classes: H\,{\sc ii} regions, ejecta nebulae and wind-blown
bubbles \citep{Chu1981}. The ejecta nebulae have clumpy or irregular
nebulosity suggesting a violent mass-loss episode and are mainly
associated with the late WNh spectral type \citep{Stock2010}, that is,
the hydrogen envelope is not completely lost. Spectroscopic studies
of nebular abundances show evidence of enrichment with material
processed in the CNO cycle in cases where the nebula is not dominated
by swept-up interstellar medium \citep[see
  e.g.,][]{Esteban1991,Esteban1992,Esteban2016,Fernandez2013}.

Since the advent of infrared (IR) astronomy, thermal dust emission has
been detected in the vicinity of hot stars and, specifically, in WR
nebulae \citep{vanBuren1988}. In particular, \citet{Marston1991}
analysed {\it IRAS} observations of the WR nebulae NGC\,2359,
NGC\,6888 and RCW\,58 around the WNE stars WR\,7, and WR\,136 and the
WN8 star WR\,40, respectively, and estimated the dust and their
nebular masses in each case from the ratio of the 60~$\mu$m to
100~$\mu$m intensities. The main conclusion of \citet{Marston1991} is
that the nebular masses in each case are too large to be composed
solely of material expelled from the central stars and so a sizeable
proportion of the nebular mass must be due to swept-up ISM
material. However, \citet{Mathis1992} also analyzed the \textit{IRAS}
data for RCW\,58 and NGC\,6888 with more sophisticated grain physics
and grain spatial distribution and concluded that the nebular mass for
RCW\,58 is in fact two orders of magnitude lower than that estimated
by \citet{Marston1991} and is therefore consistent with all the
nebular material having a stellar origen.  More recently,
\citet{Gvaramadze2010} exploited mid-IR \textit{Spitzer Space
  Telescope} data to identify compact nebulae around evolved
stars. Follow-up spectroscopic identification resulted in an increase
in the number of known Galactic LBVs and late-type WN
stars. \citet{Toala2015} analysed the mid-IR morphologies of nebulae
around WR stars using \textit{Wide-field Infrared Survey Explorer}
(\textit{WISE}) infrared images and proposed a classification scheme
that is loosely correlated with the spectral type of the central star.

Both RSG and LBV stars are known to be copious producers of dust but
its production will cease as soon as the star enters the hot WR
phase.\footnote{Dust can be also produced in the dense interaction
  region of colliding wind WC binary systems.} In RSG stars, O-rich
dust, such as silicates or Al$_2$O$_3$ forms at a few stellar radii
due to the condensation of molecules
\citep[see][]{Verhoelst2009}. Convection cells are thought to play an
important role in levitating the molecular gas above the stellar
photosphere into the upper atmosphere
\citep[][]{Chiavassa2010}. Detailed Very Large Telescope VISIR imaging
studies of the nearest RSG, Betelgeuse \citep[][]{Kervella2011}, show
that the close circumstellar environment is very clumpy, suggesting
inhomogeneous and episodic mass loss. From optical polarimetric
imaging observations of the extreme RSG VY CMa and its clumpy, dusty
circumstellar environment, \citet{Scicluna2015} derive an average
grain size of $\sim 0.5$~$\mu$m, that is, 50 times larger than grains
in the diffuse ISM, with only a small variation in grain size. The
condensation of dust in the environments of the much hotter and more
luminous LBV stars requires some sort of shielding from the star's UV
and optical emission \citep[][]{Gail2005,Kochanek2011}. It has been
proposed that dust formation in these objects is closely linked to the
eruptive state of LBVs because only very high mass-loss rates $\dot{M}
> 10^{-2.5} \mathrm{M}_\odot$~yr$^{-1}$ can provide the non-dust
optical depths required. The maximum grain size $a_\mathrm{max}$ is
related to the mass-loss rate in this scenario
\citep[][]{Kochanek2011}. Very large grains ($a_\mathrm{max} >
1$~$\mu$m) have been inferred for $\eta$~Carinae
\citep[][]{Morris2017}. Studies of very young WR nebulae, or nebulae
in low density environments far from the Galactic plane which are not
dominated by swept-up ISM material, should be able to shed some light
on the history of the dust formation process in the previous
evolutionary stage.

We have selected the iconic WR nebula M1-67 around the late-type WN
star WR\,124 (also known as 209\,BAC or Merrill's star) as an ideal
object to study the dust properties of the circumstellar nebula. It
has Galactic latitude $b = 3.31^\circ$, which places it at $z \sim
370$~pc above the Galactic plane for an assumed distance of 6.4~kpc (see
Appendix~\ref{app:distance}).
At this position the interstellar
medium density is low enough that the nebula is expected to be
composed almost entirely of material expelled from the progenitor of
WR\,124.

M\,1-67 and its central star have been objects of many
studies. WR\,124 was identified as a high-velocity late-type WN star
by \citet{Merrill1938} who suggested an association with planetary
nebulae based on the high velocity \citep[see
  also][]{Perek1967}. \citet{Minkowski1946} discovered the clumpy
H$\alpha$ emission nebula on photographic plates and it was
denominated M1-67\footnote{Object 67 in Table 1 of
  \citet{Minkowski1946}} and classified as a planetary nebula based on
appearance. Later \citet{Cohen1975} demonstrated, using primarily
extinction arguments, that Merrill's Star and M1-67 must be a
Population I WN8 star and its associated ring nebula. \citet{Solf1982}
showed that the H$\alpha$, [N\,{\sc ii}] and [S\,{\sc ii}] velocity
fields of the nebula obtained from long-slit spectra can be explained
by a nearly spherical thin shell of numerous condensations expanding
at 46~km~s$^{-1}$ with respect to a common centre. The centre of the
expansion has a heliocentric velocity of 158~km~s$^{-1}$, very similar
to the radial velocity of the central star
(175--200~km~s$^{-1}$). This is consistent with the ejecta nebula
scenario proposed by \citet{ChuTreffers1981}. A more complex dynamical
structure, consisting of an expanding hollow spherical shell ($v =
46$~km~s$^{-1}$) and a bipolar outflow ($v = 88$~km~s$^{-1}$) was
reported from a long-slit spectrographic study by
\citet{Sirianni1998}. Although M1-67 appears symmetric and almost
spherical (or hexagonal) in H$\alpha$ images, with the star at the
centre of the $\sim 110^{\prime \prime}$ diameter nebula
\citep{Grosdidier1998},\ high contrast coronographic studies
\citep{Nota1995} of the inner regions do reveal a bipolar structure,
which is also evident in \textit{Spitzer} $24$~$\mu$m images
\citep{Gvaramadze2010}. This has been attributed to anisotropic
outbursts such as are seen in LBV \citep{Nota1999}.

Spectra of M\,1-67 show strong [N\,{\sc ii}] and H$\alpha$ nebular
lines but [O\,{\sc iii}] is almost completely absent \citep[see e.g.,
][]{Esteban1991}. Abundance determinations find that N is enhanced, O
is deficient and S has normal H\,{\sc ii} region abundances. This is
further evidence that WR124 had a massive progenitor since these
abundances are an indication of material processed in the CNO
cycle and it reinforces the idea that there has been no 
significant interaction with the ISM. Moreover, the physical 
conditions in the nebula are found to be
consistent with a very low ionization level, and an electron
temperature $T_\mathrm{e} \sim 6200$ to 8200~K
\citep{Esteban1991,Fernandez2013}.

\begin{table*}
  \begin{center}
   \begin{tabular}{lcccccc}
   \hline
   \hline
   Instrument                          & Date       & Obs. ID          & PI                &    $\lambda_{\mathrm{c}}$ or $\Delta\lambda$  & Duration & Processing level \\
                                       &(yyyy-mm-dd)&                  &                   &    $(\mu$m)                                    & (s)      &                  \\
   \hline
   \hline
   {\it WISE}                          & 2010-04-11 & 2876p166\_ac51   &                   &    12 (W3), 22 (W4)                            & 8.8      &  Level 3         \\
   {\it Spitzer} MIPS                  & 2006-10-15 & 30544            &      Morris, P.   &    24                                          & 3        &  Level 2         \\
   {\it Herschel} PACS                 & 2010-04-08 & 1342194080       &    Groenewegen, M.A.T.&    70, 160                              & 2622     &  Level 3         \\
                                       &            & 1342194082       & Groenewegen, M.A.T.&    100, 160                            & 2622     &   Level 3         \\
   {\it Herschel} SPIRE                & 2010-09-21 & 1342204949       & Groenewegen, M.A.T. &    250, 350, 500                        & 911      &  Level 2         \\
   \hline
   {\it Spitzer} IRS                   & 2004-04-19 & 199              &       Houck, J.R. &  7.53--14.73 (SL1)                             & 121.9    &  Level2          \\
                                       &            &                  &                   &  14.27--21.05 (LL2)                            & 94.4    &   \\
                                       &            &                  &                   & 20.56--38.42 (LL1)                             & 94.4    &    \\
  \hline
  \end{tabular}
  \caption{Details of the IR observations used in this paper.}
  \label{tab:obsm167}
  \end{center}
\end{table*}

The distance to WR\,124 and M1-67 is not well determined and has been
a subject of discussion. Extinction studies put a lower limit to the
distance of 2.5~kpc \citep{Esteban1991}, while a spectrophotometric
distance of 8.4~kpc was determined by \citet{Hamann2006}. On the other
hand, \citet{Marchenko2010} used multi-epoch \textit{Hubble Space
  Telescope} (\textit{HST}) images to estimate a geometric distance of $3.35
\pm 0.67$~kpc to M\,1-67. The recent \textit{Gaia} Data Release 2
provides 5-parameter astrometric data for WR\,124 (positions, parallax
and proper motions), which can be combined to estimate the distance
using Bayesian methods \citep{BailerJones2018,Luri2018}. Recently, \citet{Rate2020}
presented a method to estimate distances to WR stars using the {\it Gaia} data 
including the $G$-band photometry and estimated a distance of 5.9~kpc 
for WR\,124 using a prior based on H\,{\sc ii} 
regions and dust extinction. We performed our own calculation using the 
\textit{Gaia} data and a different prior and obtained 6.4~kpc. 
In Appendix~\ref{app:distance} we describe the details of this procedure and show
that both distances are consistent within the estimated error values. Following
our result, in this paper we adopt 6.4~kpc as
the distance to WR\,124 and its nebula. At this distance, the
$110$~arcsec diameter of the nebula corresponds to a spatial size of
3.4~pc. 

\citet{Vamvatira2016} presented \textit{Herschel} imaging and
spectroscopic observations of M\,1-67. A comparison of the infrared
images with optical images shows that the dust nebula coincides with
the ionized gas, indicating that the dust and gas are
mixed. \citet{Vamvatira2016} used the 2-Dust code \citep{Ueta2003} to
model the dust emission and concluded that two populations of dust
grains were required to produce the infrared spectral energy
distribution (SED). The total mass of dust resulting from their model
is $0.22~\mathrm{M}_\odot$ of which $0.21~\mathrm{M}_\odot$
corresponds to the population of large grains with radii from 2 to
$10~\mu$m. They conclude that the progenitor of WR\,124 was a RSG with
initial mass $\sim 32~\mathrm{M}_\odot$. However, such large grain
sizes are not consistent with observations of nearby RSG
\citep{Scicluna2015} and such a low progenitor mass 
is not consistent with an LBV origin
for the nebula, such has been proposed by several
authors \citep[see, e.g., ][]{Sirianni1998,Fernandez2013}.

In this paper, we perform a detailed characterization of the dust in
M\,1-67 by analysing \textit{WISE}, \textit{Spitzer} and
\textit{Herschel} observations and using the spectral synthesis code
       {\sc Cloudy} \citep{Ferland2017} to fit simultaneously the dust
       photometric data and nebular optical data from the
       literature. This paper is organized as follows: in
       Section~\ref{sec:obs} we describe the observations used and
       detail the IR photometry. In Section~\ref{sec:work} we describe
       our photoionization model obtained with {\sc Cloudy} and the
       results of the models are given in
       Section~\ref{sec:results}. The discussion and summary are
       presented in Section~\ref{sec:disc} and \ref{sec:summ},
       respectively.

\section{OBSERVATIONS AND IR PHOTOMETRY}
\label{sec:obs}

\subsection{Observational Data}
\label{subsec:data}

We use public IR observations (images and spectroscopy) from a variety
of IR satellites. All observations were retrieved from the NASA/IPAC
Infrared Science
Archive\footnote{\url{https://irsa.ipac.caltech.edu/frontpage/}}.

We obtained IR images of M\,1-67 from {\it WISE}, {\it Spitzer} (MIPS)
and {\it Herschel} (PACS and SPIRE)\footnote{The {\it Herschel} observations 
were obtained as part of the Mass-loss of Evolved Stars (MESS) project 
\citep{Groe2011}.}. The combination of all these
observations give us a wide view of M\,1-67 covering the
12--500~$\mu$m wavelength range and enable us to construct a detailed
SED. Although {\it Spitzer} IRAC
observations are available, those images are contaminated by the
presence of a large number of stars in the field of view of M\,1-67,
which complicates the photometry extraction. Details of the individual
observations are provided in Table~\ref{tab:obsm167}.

\begin{figure*}
\centering
\includegraphics[width=0.32\textwidth]{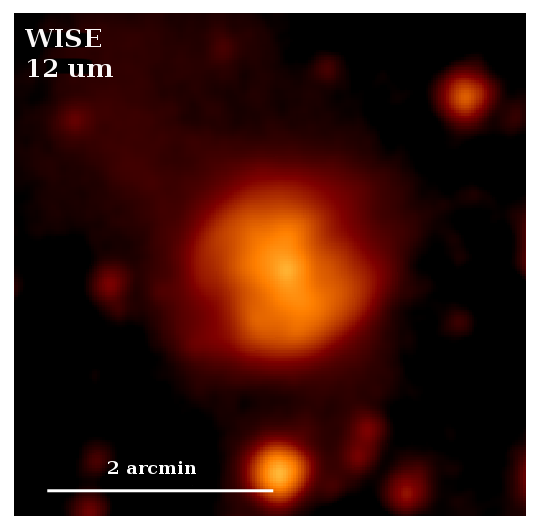}
\includegraphics[width=0.32\textwidth]{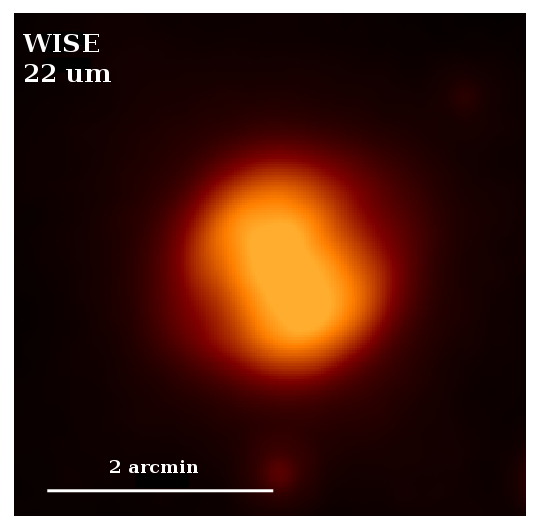}
\includegraphics[width=0.32\textwidth]{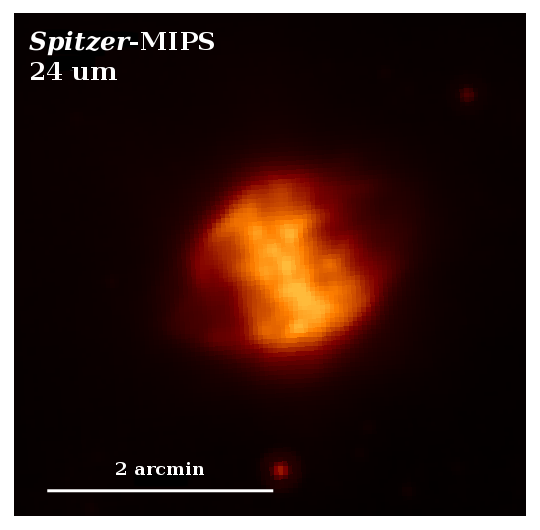}\\
\includegraphics[width=0.32\textwidth]{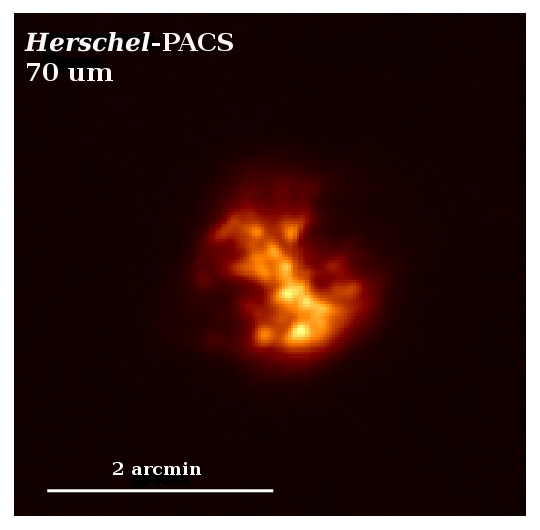}
\includegraphics[width=0.32\textwidth]{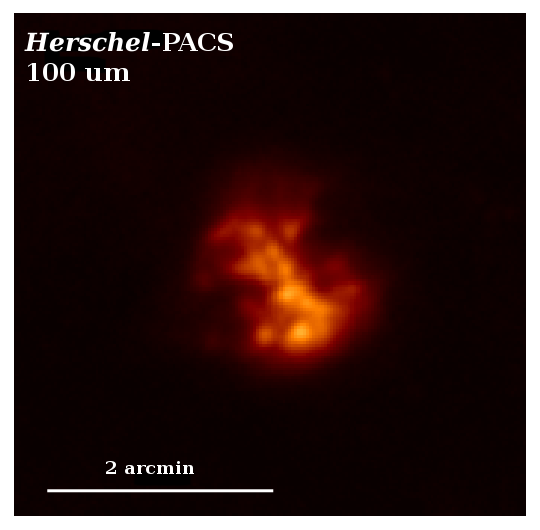}
\includegraphics[width=0.32\textwidth]{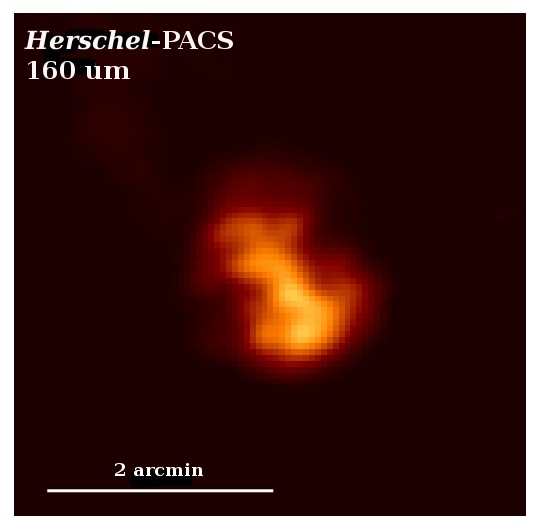}\\
\includegraphics[width=0.32\textwidth]{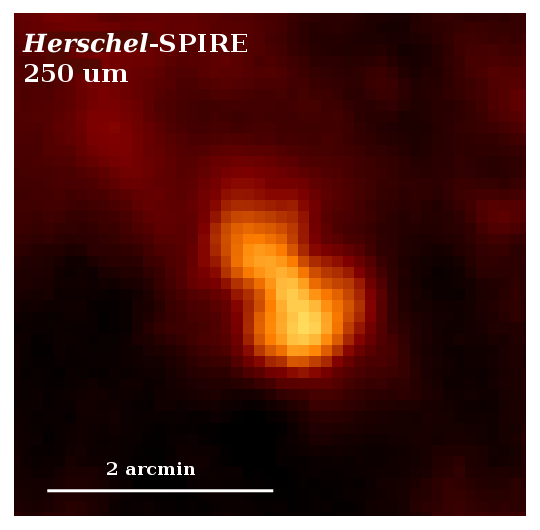}
\includegraphics[width=0.32\textwidth]{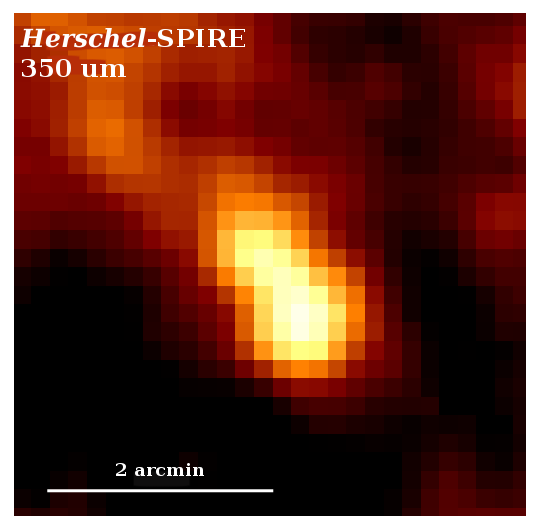}
\includegraphics[width=0.32\textwidth]{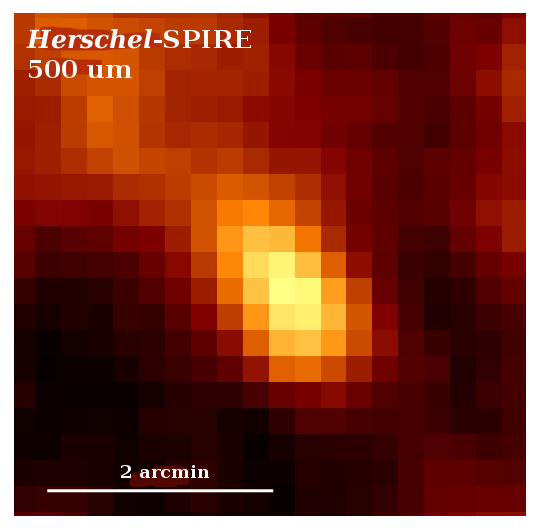}
\caption{Images of the nebula M\,1-67 obtained by \textit{WISE},
  \textit{Spitzer} and \textit{Herschel}. The different bands are
  centered at 12, 22, 24, 70, 100, 160, 250, 350 and 500~$\mu$m. North
  is up and east to the left.}
\label{fig:m167mw}
\end{figure*}

Figure~\ref{fig:m167mw} shows all of the IR images obtained from
\textit{WISE}, \textit{Spitzer} (MIPS) and \textit{Herschel} (PACS and
SPIRE). These IR images clearly show a bipolar morphology \citep[see,
  e.g.,][]{Sirianni1998}, in contrast to the symmetrical nebula seen
in deep {\it HST} WFPC2 H$\alpha$ image of M\,1-67
\citep[][]{Grosdidier1998}. This is appreciated to some extent even in
the long wavelength \textit{Herschel} SPIRE images (bottom row of
Fig.~\ref{fig:m167mw}). Some extended emission can be seen in the
\textit{WISE} 12~$\mu$m and the {\it Herschel} SPIRE images, which is
very likely to be contribution from material in the line of sight
rather than material at the location of M\,1-67, given its position
above the Galactic plane.

We also obtained \textit{Spitzer} Infrared Spectrograph (IRS)
low-resolution data, which cover a spectral range of
7--38~$\mu$m. These observations were processed using the CUbe Builder
for IRS Spectral Maps \citep[{\sc cubism};][]{Smith2007}. 

\subsection{IR photometry}
\label{subsec:phot}

\begin{figure}
\centering
\includegraphics[width=\linewidth]{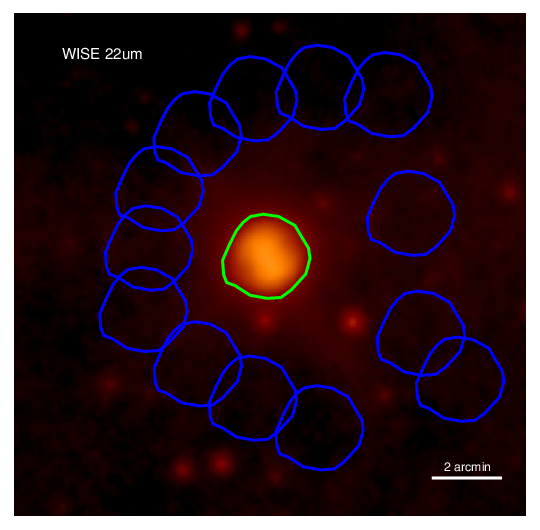}
\caption{Examples of background selection regions around M\,1-67. The
  image shows the \textit{WISE} band~4. The region used for extracting
  the flux from M\,1-67 is shown with a green contour and the
  background regions are shown in blue.}
\label{fig:m167b22}
\end{figure}

\begin{figure}
   \centering
\includegraphics[width=\linewidth]{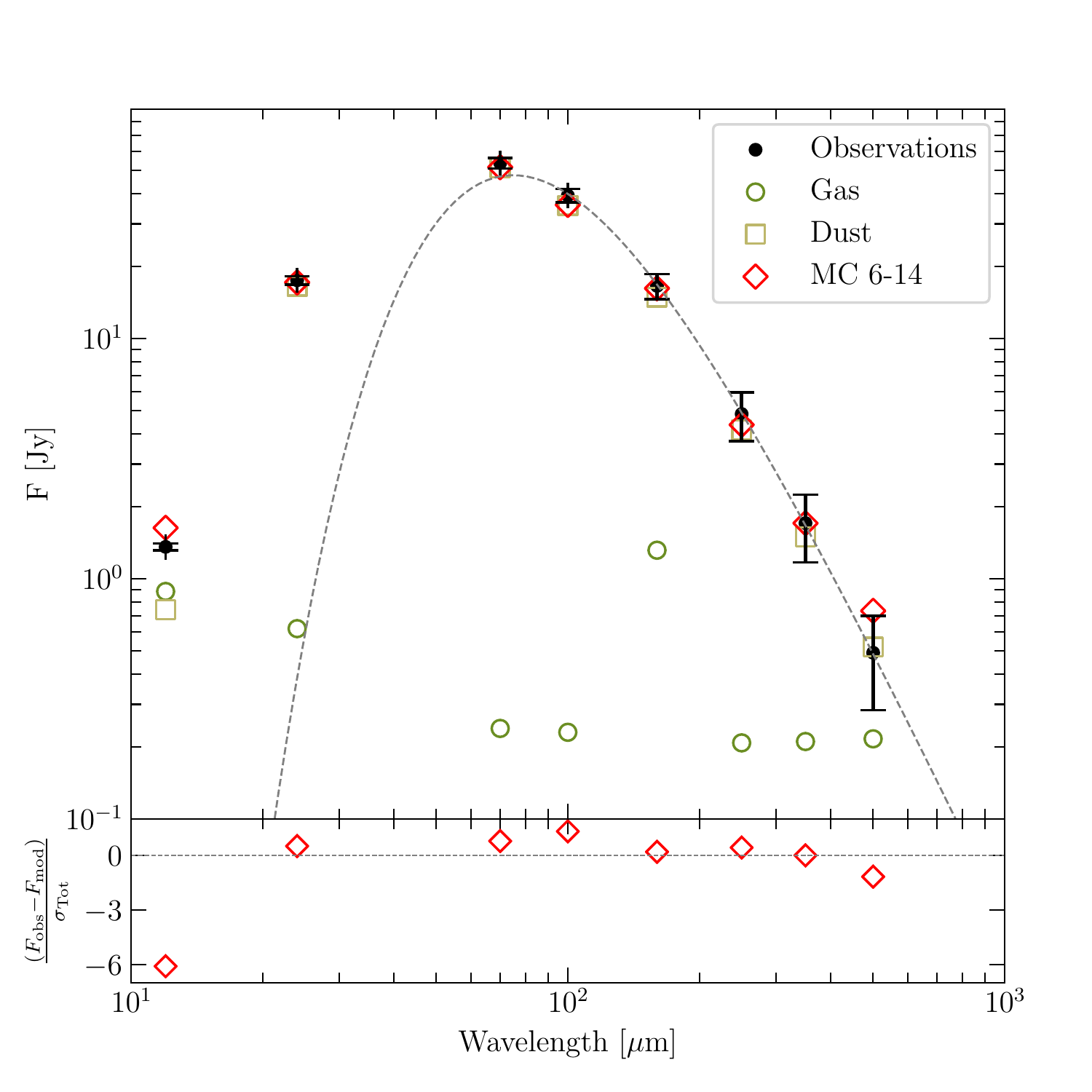}
\caption{SED obtained from the IR observations of M\,1-67: (black) dots
  (see Table~\ref{tab:fot}). The error bars take into account
  uncertainties associated with the instrument calibration and the
  background subtraction process. The synthetic SED obtained from our
  best model, MC\,06-14, is also shown with empty red diamonds (see
  Section~4.2 for details). The contribution to the synthetic photometry
  from the ionized gas and dust are also shown. The residual is defined 
  as $(F_\mathrm{obs}-F_\mathrm{mod})/\sigma_\mathrm{Tot}$. 
  The dashed line 
  show the best fit (for
  $\lambda \geq 100 \mu$m) to the \textit{Herschel} PACS and SPIRE
  photometry for a modified black body model, see Section~\ref{ssec:mbb} for details}. 
\label{fig:SED}
\end{figure}

In order to extract a SED from M\,1-67, we defined a region that
encompasses the nebular emission for each individual IR image. We then
integrated the flux density for each region so defined. A background
correction was applied to the flux value from each image. The
background is not uniform and is different at each wavelength. Thus,
the selected background regions for each IR image are not the same as
those for a different image. The backgrounds were selected by eye and
a set of $n$-background regions was obtained for each image. We
subtracted each background from the nebular flux density in order to
obtain $n$ background-subtracted fluxes for each IR wavelength
image. The mean and standard deviation of these values then give us
the final flux value and measurement uncertainty
($\sigma_{\mathrm{back}}$), respectively. Figure~\ref{fig:m167b22}
shows an example of the background selection process.

To calculate the total error from each IR band we add together the
different contributions, as well as the uncertainty due to the
background subtraction. The calibration uncertainty
($\sigma_{\mathrm{cal}}$) from each instrument has to be taken into
account, too. Then, the total error is given by
\begin{equation}
\sigma_{\mathrm{ Tot}} = \sqrt{
  \sigma_{ \mathrm{back} }^2 + \sigma_{ \mathrm{cal} }^2}.
\end{equation}
For the \textit{WISE} data, the calibration uncertainty is 4.5\% and
5.7\% from the W3 and W4 bands, respectively. The \textit{Herschel}
data calibration uncertainty is 5\% for the PACS and 7\% for the SPIRE
instruments, while that for the \textit{Spitzer} MIPS is 4\%.

We note that the photometry of M\,1-67 extracted from the {\it
    WISE} W3 band has a significant contribution from the central star,  WR\,124. We
  carefully subtracted the contribution from WR\,124 by comparing the
  PSF of stars in the field. This produced a clean photometric measure
  of the W3 12~$\mu$m band.
 
In Table~\ref{tab:fot} we present the final flux estimates for each IR
image of M\,1-67 with their corresponding uncertainties
($\sigma_{\mathrm{Tot}}$) associated with the instrument calibration
and the background subtraction process. The final IR SED of M\,1-67 is
presented in Figure~\ref{fig:SED}.

\begin{table}
  \begin{center}
     \begin{tabular}{cccccc}
     \hline
     \hline
     Instrument  &$\lambda_\mathrm{c}$ & Flux              & $\sigma_{\mathrm{f}}$ & $\sigma_{\mathrm{cal}}$ & $\sigma_{\mathrm{Tot}}$ \\
                          &$[\mu$m]  & [Jy]              &                       &                         &                         \\
     \hline
     {\it WISE}            & 12        &  1.36             & 0.05                  & 0.06                    & 0.08     \\
                           & 22        & 17.91             & 0.04                  &  1.02                   & 1.80     \\
     {\it Spitzer} MIPS         & 24        & 17.47             & 0.08                  &  0.70                   & 0.7      \\
     {\it Herschel} PACS   & 70        & 53.72             & 0.25                  &  2.69                   & 2.70     \\
                           & 100       & 39.34             & 1.62                  &  1.97                   & 2.55    \\
                           & 160       & 16.55             & 1.79                  &  0.83                   & 1.98    \\
     {\it Herschel} SPIRE  & 250       & 4.85              & 1.06                  &  0.34                   &  1.11   \\
                           & 350       & 1.70              & 0.53                  &  0.12                   &  0.53   \\
                           & 500       & 0.49              & 0.21                  &  0.03                   & 0.21    \\
     \hline
     \hline
     \end{tabular}
     \caption{Flux densities and uncertainties of different IR band
       observations of M\,1-67. $\sigma_{\mathrm{f}}$ is the
       uncertainty associated with background inhomogeneity,
       $\sigma_{\mathrm{cal}}$ is the uncertainty associated with the
       instrument calibration of each observation and
       $\sigma_{\mathrm{Tot}}$ is the sum of these uncertainties.}
     \label{tab:fot}
  \end{center}
\end{table}

\subsection{Modified Black Body Model}
\label{ssec:mbb}
A modified black body fit (MBB) to the dust emission in the far infrared 
($\lambda \geqslant 100 \mu$m) is a useful tool for estimating the 
dust mass and mean dust temperature of the coolest, i.e., the largest 
dust grains in M\,1-67.
The underlying assumption is that the dust is optically thin 
in local thermodynamic equilibrium (LTE) and that a
single temperature $T_\mathrm{D}$ and emissivity index $\beta$ are
representative of all the dust
grains. The MBB fit to the observed flux densities is
\begin{equation}
F_\nu = M_\mathrm{D} \kappa_{\nu}  \frac{B_\nu (T_\mathrm{D})}{d^2},
\label{eq:dfunc}
\end{equation}
\noindent where $d$ is the distance to the nebula, $B_\nu(T_\mathrm{D})$ 
is the Planck function and $M_\mathrm{D}$ is the total dust mass. 
The dust emissivity
$\kappa_\nu$ is an exponential function of frequency parametrized by the
emissivity index $\beta$, that is
\begin{equation}
 \kappa_\nu = \kappa_{\nu 0} \left( \frac{\nu}{\nu_0} \right)^\beta \ .
\label{eq:knorm}
\end{equation}
The dust emissivity normalization for $R_V = 3.1$ Milky Way dust is 
$\kappa_{\nu_0} = 1.92$~cm$^2$g$^{-1}$ at the 
reference wavelength $\nu_0 = 350~\mu$m
\citep{Draine2003}.

The MBB fit returns the dust temperature $T_\mathrm{D}$ and the
emissivity index $\beta$ of the grains responsible for the emission at
wavelengths $\lambda \ge 100 \mu$m. If $d$ is known, then $M_\mathrm{D}$ 
provides the normalization of the model.

The fitting procedure minimizes a $\chi^2$ function which fits the
model described in Equation~\ref{eq:dfunc} to the observed SED (see
Fig.~\ref{fig:SED} and Table~\ref{tab:fot}). 3000 fluxes are drawn
from a randomly sampled normal distribution centred on the measured
values with standard deviation equal to the uncertainty in each
photometry measurement. We use only the \textit{Herschel} PACS and
SPIRE photometry for the fit, since these measurements have the least
contamination from emission lines.

The total dust mass resulting from this procedure is $M_\mathrm{D} =
0.36 \pm 0.09$~M$_\odot$ and the characteristic dust temperature is
$T_\mathrm{D} = 37.52 \pm 6.51$~K. The emissivity index returned by
the fit is $\beta = 2.02 \pm 0.38$. 
The obtained $\beta$ value is sufficiently close
to the spectral index of the underlying theoretical dust model,
$\beta=2$, suggesting that the emission for $\lambda \ge 100~\mu$m
comes from dust in a very narrow temperature range, i.e., 
very likely a dust population with a small range of grain sizes. 
This MBB fit is shown in Figure~\ref{fig:SED} in comparison with the
observed SED. This figure shows that 
the complete SED is too broad to be fit by a single population 
of grains. We conclude that a detailed model should include at least 
two populations of grains, as has been suggested from previous IR studies 
of WR nebulae \citep[e.g.,][]{Mathis1992}.

It is important to note that the MBB fit gives us no information on 
the properties of the small grains, which absorb most of the stellar 
radiation and are responsible for the majority of the mid-infrared 
emission from the nebula. 
Nevertheless, the value of $T_\mathrm{D}$ estimated in this section 
will be a useful guide to restrict the large grain temperature in our 
detailed, self-consistent photoionization 
model of M\,1-67 (see below).

\section{Detailed modelling of M\,1-67}
\label{sec:work}

Our aim is to produce a model of M\,1-67 that can simultaneously
explain both the nebular optical properties and its infrared
photometry and spectroscopy. For our reference optical observations,
we use the high spatial resolution spectroscopic results presented by
\citet{Esteban1991}. We chose these observations rather than the more
recent integral field spectroscopy observations of
\citet{Fernandez2013} because the slit positions of
\citet{Esteban1991} give spectra that are more representative of
average conditions in the nebula. Furthermore, we assume that M\,1-67
can be modeled as a purely photoionization nebula since there are no
apparent signs of interaction between a fast wind and the clumpy
nebula material \citep{Fernandez2013}. The kinematics of the clumps
suggests an ejecta origen rather than the breakup of a swept-up shell
due to instabilities \citep{Solf1982,Grosdidier1998,Toala2011}.

The nebular material consists of gas and dust, both of which interact
with the UV flux from the central star and so cannot be treated
separately. The spectral synthesis and plasma simulation code \textsc{Cloudy}
\citep[][]{Ferland2017} is the ideal tool to investigate this
interaction, coupled with the \textsc{pyCloudy} library \citep{Morisset2014} to
obtain the fluxes and line intensities through apertures corresponding
to the different instruments and observations we wish to model.

\textsc{Cloudy} uses as input parameters: (i) the stellar properties
(luminosity and spectral shape), (ii) the nebular properties
(abundances and radial density profile) and (iii) dust properties
(species, size distributions and dust-to-gas ratio). Calculations were
performed with version 17.01 of \textsc{Cloudy}, last described by
\citet{Ferland2017}.

\subsection{Stellar atmosphere}
\label{ssec:stell}

The spectra of WN-type WR stars are the result of non-LTE radiative
transfer through expanding atmospheres including line blanketing by
iron and iron-group elements \citep{Grafener2002} and, as such, the
spectral shape departs strongly from that of a black body spectrum. In
particular, the absorption of UV photons drives powerful stellar
winds, which have a ratio of mechanical to radiation momentum much larger
than unity due to multiple scattering.  Detailed synthetic spectra
should be used when modelling the observable properties of WR nebulae
\citep[see][]{Reyesperez2015} and in this paper we use the Potsdam
Wolf-Rayet (PoWR) model atmospheres \citep{Hamann2006}, which have
been developed over many years and include effects such as
line-blanketing and clumping in the
wind.\footnote{\url{http://www.astro.physik.uni-potsdam.de/~wrh/PoWR/powrgrid1.php}}

\begin{figure}
\centering
\includegraphics[width=\linewidth]{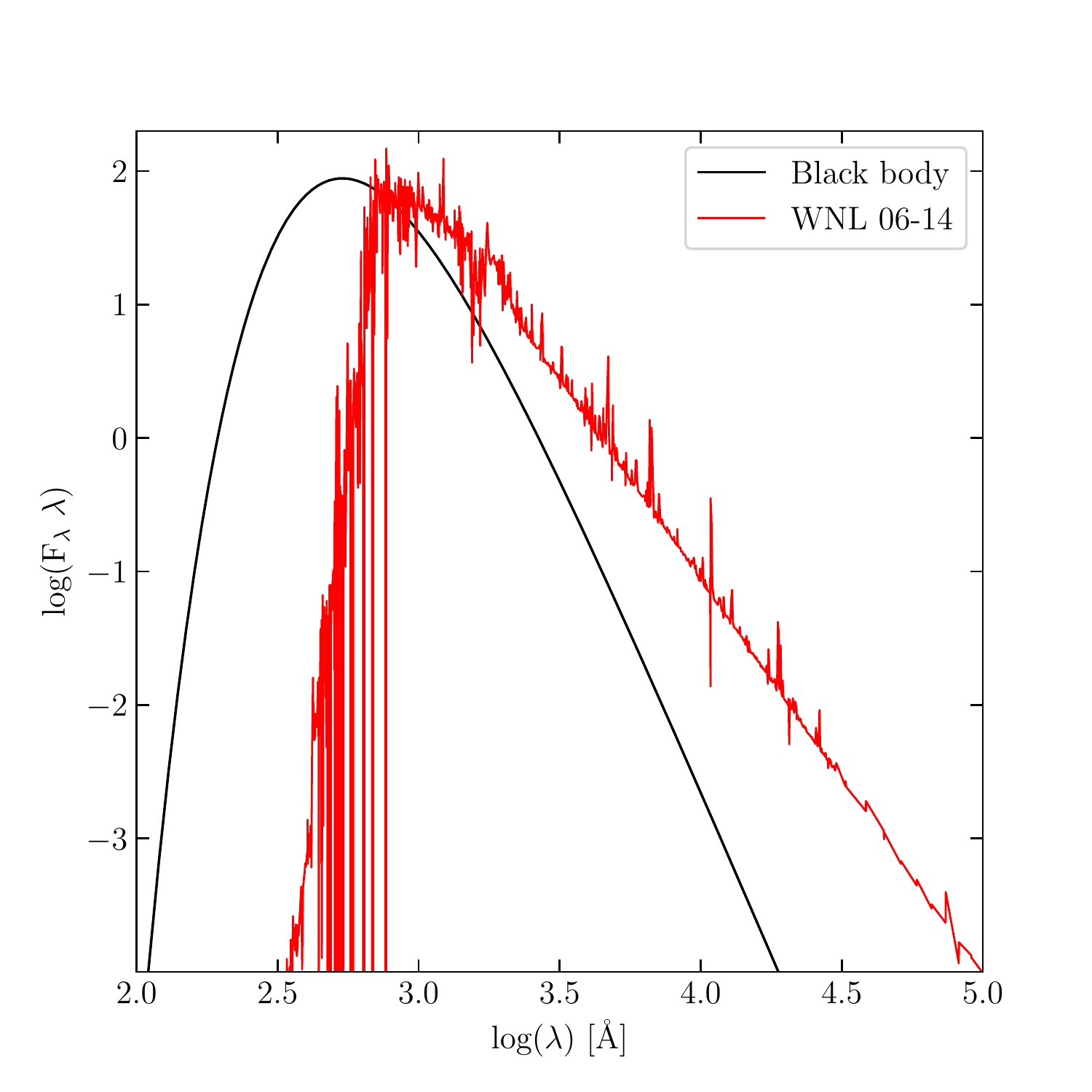}
\caption{Comparison between black body emission and
  PoWR stellar atmosphere model. Solid (red) line: best-fit model to
  WR\,124 determined by \citet{Hamann2006} corresponding to model
  WNL\,06-14 with $T_{\mathrm{eff}}=44.7$\,kK. The black line
  represents the black body spectrum with the same effective
  temperature.}
\label{fig:pwbb} 
\end{figure}

WR\,124 was given the spectral classification WN8h on the basis of its
optical spectrum \citep{Smith1969}. \citet{Hamann2006} modelled the
atmosphere of this star with the PoWR code and the best fit in
$R_\mathrm{t}$-$T_\mathrm{eff}$ space to the {\it IUE} flux and narrow-band
visual photometry is model 06-14 from the WNL-H20 grid of models. Here,
$R_\mathrm{t}$ is the transformed radius, related to the mass-loss
rate $\dot{M}$ by
\begin{equation}
  R_\mathrm{t} = R_* \left\{  \frac{v_\infty}{2500 \mathrm{km\,s}^{-1}} \bigg/ \frac{\dot{M}\sqrt{D}}{10^{-4} M_\odot \mathrm{yr}^{-1}}  \right\}
    \label{eq:trlaw}
\end{equation}
where $R_*$ is the stellar radius, $v_\infty$ is the stellar wind
velocity and $D$ is a clumping factor. Stellar luminosity, temperature
and radius are related through the Stefan-Boltzmann law. The radius,
luminosity and mass-loss rate can be scaled with distance using the
transformation law Equation~\ref{eq:trlaw}. Details of the WNL\,06-14
PoWR grid model are listed in Table~\ref{tab:powr}. In
Figure~\ref{fig:pwbb} we show the comparison between the synthetic
spectrum WNL\,06-14 obtained with the PoWR model atmosphere code and
black body radiation with the same effective temperature. The model atmosphere
is extremely deficient in EUV photons compared to the black body
spectrum due to photon reprocessing in the stellar
wind driving region. We rescaled the luminosity of WR\,124 calculated by \citet{Hamann2019} 
to our adopted distance of 6.4~kpc,
and obtain a luminosity
of $L = 10^{5.76}\mathrm{L}_{\odot}$ for WR\,124. The WN8 class of
stars is among the most luminous stars known.

\begin{table}
  \begin{center}
     \begin{tabular}{lc}
     \hline
     \hline
     Model                                           &  WNL\,06-14           \\
     \hline
     $X_{\mathrm{H}}$ [\%]                             & 20                  \\
     $T_\mathrm{eff}$ [kK]                             & 44.7                \\
     $\log_{10}R_{\mathrm{t}}$  [R$_{\odot}$]          & 0.7                 \\
     $\log_{10}Q_{\mathrm{H}}$  [s$^{-1}$]             & 48.85               \\
     $\log_{10}Q_{\mathrm{He}}$ [s$^{-1}$]             & 44.91               \\
     $D_{\infty}$                                      &  4                  \\
     \hline
     \end{tabular}
     \caption{Parameters of PoWR Model WNL\,06-14, which  was found by \citet{Hamann2006} to be the best fit
       to the stellar spectrum of WR\,124.}
     \label{tab:powr}
  \end{center}
\end{table}

\subsection{Nebular properties}
\label{ssec:nprop}

In H$\alpha$ images M\,1-67 has a clumpy appearance, while kinematic
studies suggest it is comprised of an expanding hollow shell of
clumps. \citet{Grosdidier1998} found that the radial H$\alpha$
surface-brightness distribution can be reproduced by a radial
power-law density distribution of the form
\begin{equation}
n_\mathrm{e}(r) = n_{0} \left(\frac{r_\mathrm{in}}{r}\right)^{\alpha},
\label{eq:plaw}
\end{equation}
with power-law index $\alpha = 0.8$. Furthermore, the nebula must be
density bounded with an external cutoff radius because H$\alpha$
images show that the nebula is completely
ionized~\citep{Grosdidier1998}. \citet{Fernandez2013} report electron densities of
$n_{\mathrm{e}} \sim 1500$~cm$^{-3}$ in the central regions of the
nebula and $n_{\mathrm{e}} \sim 650$~cm$^{-3}$ in the outer regions
with a symmetric gradient. \citet{Sirianni1998} calculate the 
electron densities at 270 positions in M\,1-67 and find densities as 
low as 150~cm$^{-3}$ in external parts of the nebula and some central 
positions with densities as high as 2400~cm$^{-3}$. \citet{Vamvatira2016} make the
approximation that the nebula is a spherical shell with inner radius
$r_\mathrm{in} = 45$~arcsec \citep{Solf1982} and outer radius
$r_\mathrm{out} = 60$~arcsec \citep{Grosdidier1998}. In our models, we
adopt a density distribution similar to Equation~\ref{eq:plaw} and
vary $n_{0}$, $r_\mathrm{in}$ and $\alpha$. We also consider multiple
shells and take in to account the bipolar appearance of the dust 
emission evident in Figure~\ref{fig:m167mw}.

\citet{Esteban1991} derive chemical abundances for 4 positions in
M\,1-67, while \citet{Fernandez2013} determined abundances for 8
different regions; both studies agree that nitrogen is enriched while
oxygen is deficient in the nebula as a result of processing in the CNO
cycle. Their results are summarized in Table~\ref{tab:abun}. We note, 
however, that the derived chemical abundances from these studies 
vary with position in M\,1-67.

\begin{table}
  \begin{center}
     \begin{tabular}{ccc}
     \hline
     \hline
     & \citet{Esteban1991}$^{1}$ & \citet{Fernandez2013}$^{2}$  \\  
     \hline
O    & 8.12           & 7.70  \\
S    & 6.96           & 6.40   \\
N    &  8.54          & 8.21  \\
     \hline
     \end{tabular}
     \caption[]{Chemical abundances in M\,1-67
       shown in 12+log$_{10}X$ notation. $^{1}$ Region A. $^{2}$ Regions 5 and 6.}
     \label{tab:abun}
  \end{center}
\end{table}

\begin{figure}
\centering
\includegraphics[width=\linewidth]{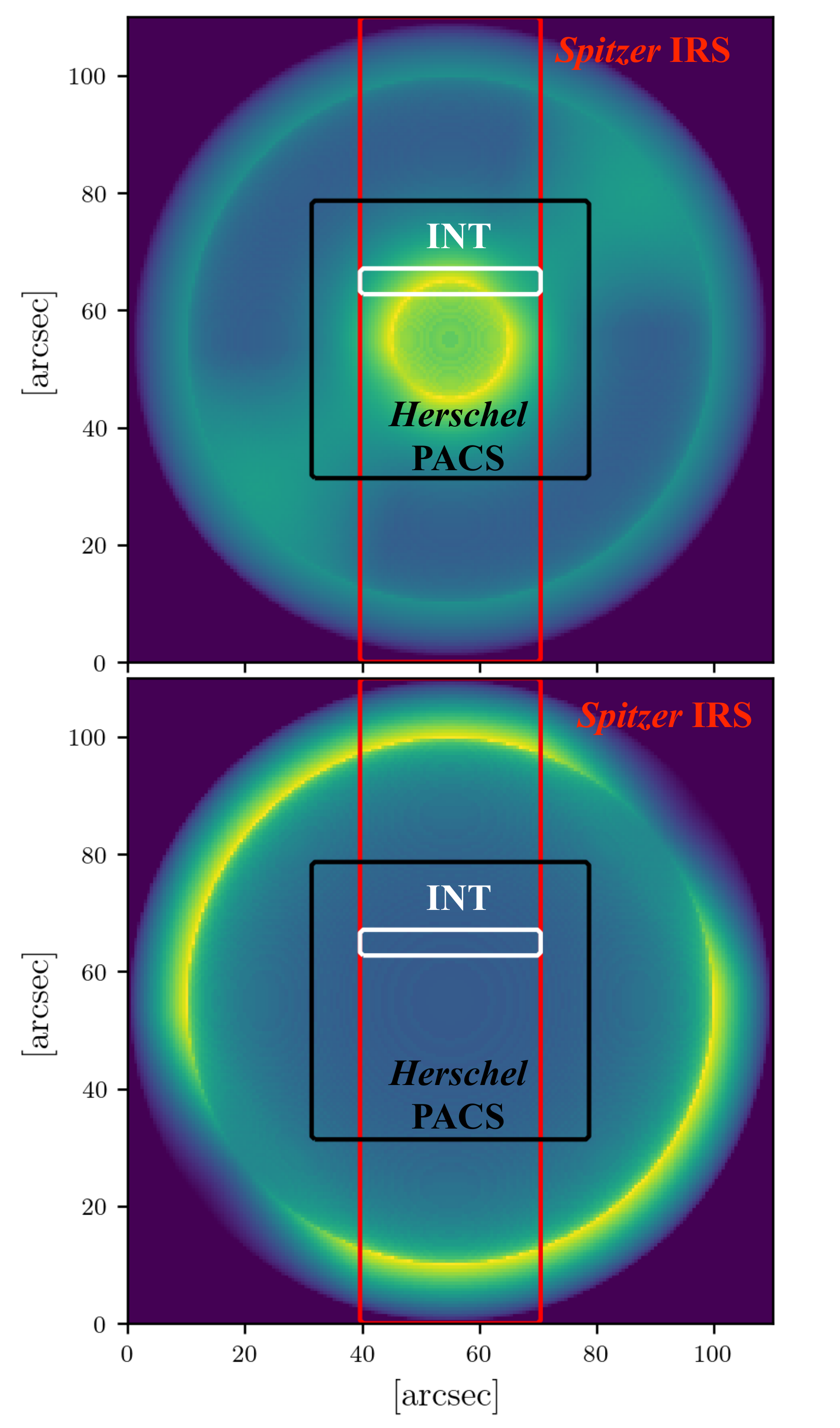}
\caption{Synthetic emission maps obtained from our best model 
MC\,6-14 calculated by \textsc{pyCloudy}. Top panel: H$\beta$ emissivity 
on a logarithmic scale. Bottom panel: {\it Herschel} PACS 70~$\mu$m. 
The rectangular regions show the relative
    positions and sizes of apertures representing region A
    of the INT slit from \citet{Esteban1991} (white), footprint of the 
    \textit{Herschel}-PACS spectrometer (blue),
    and the slit of \textit{Spitzer} IRS observations (red).}
\label{fig:pycl} 
\end{figure}

\subsection{Dust properties}
\label{ssec:dprop}

The photospheres of evolved massive stars are oxygen-rich, rather than
carbon-rich, due to nucleosynthesis by the CNO cycle in the core,
which favours the production of $^{14}$N at the expense of $^{12}$C. The
dust formed in RSG and LBV should therefore be primarily
calcium-magnesium-iron silicates, such as olivine (MgSiFeO$_4$), and
metal oxides, such as alumina (Al$_2$O$_3$) \citep[see,
  e.g.,][]{Gail2005,Cherchneff2013}. We examined the publicly available
{\it Spitzer} IRS spectra to search for grain emission features but found no
evidence for crystalline silicates at their known wavelengths 6, 11.3,
19, 23, 27, 33~$\mu$m \citep{Henning2010}. Accordingly, we assume the
grains are composed of amorphous astronomical silicate material
\citep{Greenberg1996}, specifically olivine (MgFeSiO$_4$), which is the default
silicate grain option in \textsc{Cloudy}.

The grains are spherical and are resolved into size bins described by a power-law
distribution $N(a) \propto a^{-3.5}$ \citep[MRN size
  distribution;][]{Mathis1977}, with typically 10 size bins between
$a_\mathrm{min}$ and $a_\mathrm{max}$ where the minimum and maximum
grain sizes are free parameters. The dust-to-gas mass ratio ($D/G$) in
\textsc{Cloudy} can be adjusted by the user but the default value defines 
the Si abundance in the silicate dust to be $3.28\times10^{-5}$, equivalent 
to the Solar value, and the mass number per dust molecule is 172 
\citep[e.g.,][]{Weingartner2001}.

\begin{table*}
  \begin{center}
     \begin{tabular}{lccccccc}
     \hline
     \hline
Parameter                                & MA\,6-14   & \multicolumn{2}{c}{MB\,6-14} & \multicolumn{4}{c}{MC\,6-14} \\  
                                         &             & Inner shell & Outer shell &\multicolumn{2}{c}{Inner Shell} & \multicolumn{2}{c}{Outer Shell}\\ 
 & & & & Zone 1 & Zone 2 & Zone 3 & Zone 4 \\
\hline
Distance [kpc]                           & 6.4         & 6.4         & 6.4    & 6.4         & 6.4         & 6.4         &  6.4    \\
$\log_{10}(L_{\star}/\mathrm{L_\odot})$   & 5.76        & 5.76       & 5.76   & 5.76        &  5.76       & 5.76        &  5.76    \\
$c(\mathrm{H}\beta)$                     & 1.9         & 1.9         & 1.9    & 1.9         &  1.9        & 1.9         &  1.9    \\
Inner radius["]                          & 10          & 10          & 45     & 10          &  10         & 45          &  45     \\
Outer radius["]                          & 55          & 45          & 55     & 45          &  45         & 55          &  55     \\
     \hline
$n_{\mathrm{0}}$\,[cm$^{-3}$]            & 2100         & 2500       & 600    & 2500        & 900         & 600         & 666\\  
$\alpha$                                 &  1.4         & 2.0        & 2.0    & 2.0         & 0.2         & 2.0         & 2.0\\  
Filling Factor                           &  0.05        & 0.05       & 0.037  & 0.05        & 0.02        & 0.0365      & 0.005 \\
\hline
$a_\mathrm{small}$ [$\mu$m]              & 0.005--0.25$^{\mathrm{(a)}}$  & - & 0.005--0.05$^{\mathrm{(a)}}$ & -     & -     & 0.005--0.05$^{\mathrm{(a)}}$       & -  \\
$a_\mathrm{big}$ [$\mu$m]                & -                             & - & 0.9$^{\mathrm{(b)}}$         & -     & -     & 0.9$^{\mathrm{(b)}}$               & -  \\
$B/S$                                    & -                             & - & 20                           & -     & -     & 20                                 & -  \\
$D/G$                                    & 4.03$\times 10^{-3}$         & - & 3.87$\times 10^{-2}$          & -     & -     & 6.33$\times 10^{-2}$               & -  \\
\hline
Gas mass [M$_{\odot}$]                   &  11.87                 & 5.48        &  5.78                 & 3.29  & 2.19  & 3.41 & 0.35  \\
Dust mass [M$_{\odot}$]                  &  0.05                  &  -          &  0.22                 & -     & -     & 0.22 & -     \\
\hline
\hline
\end{tabular}
\caption{Input parameters for the model sets discussed in the
  text.$^{\mathrm{(a)}}$ Population of grains with a standard MRN grain size
  distribution. $^{\mathrm{(b)}}$ Population of grains represented by a single size.}
\label{tab:input}
\end{center}
\end{table*}

\subsection{Synthetic spectra and photometry}

For a given set of input parameters, \textsc{Cloudy} calculates the
ionization and thermal equilibrium solution and computes the continuum
and line volume emissivities as a function of radius and depth for the
full spectral range from X-rays to radio wavelengths. We used the
\textsc{pyCloudy} library \citep[see][]{Morisset2006} to produce
surface brightness maps from the spherically symmetric emissivity
results of our models. We extracted simulated optical spectra from a
synthetic aperture with the same angular size and position as slit A
of the Isaac Newton Telescope (INT) observations presented by
\citet{Esteban1991} (see Figure~\ref{fig:pycl}).

We also simulated the broadband infrared photometry corresponding
to the {\it WISE}, {\it Spitzer} and {\it Herschel} observations shown in Figure~\ref{fig:m167mw} and
Table~\ref{tab:fot}. The continuum emission is integrated over each
photometer band weighted by the corresponding transmission curve\footnote{\url{http://svo2.cab.inta-csic.es/theory/fps/index.php?mode=browse}}.

\subsection{Constraints on the models}
  \label{ssec:constr}

In order to assess if a model is a good fit to the observations, we
need to define some constraints. We selected the following:
\begin{enumerate}

\item the total observed H$\alpha$ nebular flux. This was taken from
  \citet{Grosdidier1998};

\item the H$\beta$ surface brightness $F(\mathrm{H}\beta)$ otained
  from the INT slit~A aperture taken from \citet{Esteban1991};

\item the gas electron temperature $T_\mathrm{e}$ derived from the [N\,{\sc
    ii}]\,$\lambda\lambda$\,5755/6584\, ratio
  from the INT slit~A aperture taken from \citet{Esteban1991};

\item the shape and flux of the IR photometry obtained from all images
  shown in Figure~1, illustrated in Figure~3 and listed in
  Table~\ref{tab:fot}.

\end{enumerate}

We do not attempt to fine-tune the chemical abundances of our models
and we keep them fixed at the values reported by
\citet{Esteban1991}. This may lead to differences between observed and
model emission-line intensities for some elements, but line ratios
should not be affected. We note that derived conditions across M\,1-67
are not uniform due to the clumpiness of the nebula 
\citep{Esteban1991,Fernandez2013}, but we do not model discrete clumps in  the present work.

\section{Results}
\label{sec:results}

In this section, we describe the results of a series of models varying
the density distribution and the dust properties, leading to a
self-consistent description of the nebular optical and IR properties
of M\,1-67.

\subsection{Gas Density Distribution}
\label{ssec:nebul}

We begin by investigating the gas density distribution. We ran a set of
\textsc{Cloudy} models using the PoWR 06-14 stellar atmosphere model
\citep{Hamann2006} as input and including pure silicate grains with an
MRN power-law size distribution represented by 10 size bins in the
range 0.005\,$\mu$m to 0.25\,$\mu$m, which is the standard ISM size
distribution. We fixed the outer radius of the nebula,
$r_{\mathrm{out}} = 55$\,arcsec~\citep{Grosdidier1998} and allowed the
inner radius $r_\mathrm{in}$ to vary. The density in the shell obeys a
power-law distribution given by Equation~\ref{eq:plaw}, where the
power-law index and the density at the inner radius are to be
determined and we assume spherical symmetry. 
Using the constraints of the observed total $F$(H$\alpha$)
flux and the $F$(H$\beta$) surface brightness and the electron
temperature $T_\mathrm{e}$, we varied the inner radius
$r_\mathrm{in}$, the density at the inner radius $n_{0}$ and the
power-law index $\alpha$ until the model estimates of the observed
constraint parameters agreed within 25\% for each value. The best
model, labeled as MA\,6-14, resulted from the parameter values
$n_0=2100$~cm$^{-3}$ and $\alpha=1.4$,
with inner and outer radii
of 10 and 55~arcsec, respectively.

\begin{table}
\begin{center}
\setlength{\tabcolsep}{0.5\tabcolsep} 
     \begin{tabular}{llcccccc}
     \hline
     \hline
Instrument                & Parameter               &Observations      &  MA\,6-14  & MB\,6-14  & MC\,6-14  \\
\hline 
{\it HST}$^\mathrm{(a)}$  &  log$_{10}F$(H$\alpha$) & $-$10.55$^{(1)}$ & $-$10.56 & $-$10.55  &  $-$10.55   \\
INT$^\mathrm{(b)}$        & log$_{10}F$(H$\beta$)   & $-$13.01$^{(2)}$ & $-$13.03 & $-$13.07  &  $-$13.105  \\
     \hline
     \end{tabular}
     \caption{Emission lines used to constrain our models: 
     $^\mathrm{(a)}$~Total H$\alpha$ flux, $^\mathrm{(b)}$~H$\beta$ surface brightness 
     through INT slit A. The errors from the observational
       measurements are statistical, of order
       $\sim5\%$. $^{(1)}$~\citet{Grosdidier1998}.$^{(2)}$~\citet{Esteban1991}. }
        \label{tab:linesfit}
\end{center}
\end{table}

Input details of model MA\,6-14 are listed in column~2 of
Table~\ref{tab:input}. The output values of the constraint
parameters for this model are given in
Table~\ref{tab:linesfit}. Additionally, in Table~\ref{tab:linesout},
model predicted emission-line intensities are compared to observed
values at optical and IR wavelengths from different apertures. We note
that although the low-ionization lines of ions such as [N\,{\sc ii}]
are well reproduced by the model, higher ionization lines, for
example, [S\,{\sc iii}] are not well-modeled. We attribute this to the
stellar atmosphere model, which lacks photons
capable of photoionizing S to S$^{++}$. We also note that this model
produces a reasonable amount of emission in the [C\,{\sc ii}]
$\lambda$\,158\,$\mu$m line. \citet{Vamvatira2016} interpreted this
emission as being produced in a photon-dominated region (PDR) but 
our models are fully ionized and density bounded.

\begin{table}
\begin{center}
\setlength{\tabcolsep}{0.5\tabcolsep}
     \begin{tabular}{llcccccc}
     \hline
     \hline
Instrument  & Line                                  & Observed  & MA\,6-14        & MB\,6-14 & MC\,6-14\\
\hline
INT$^{(1)}$& H$\beta$                               &100.0      & 100.0            & 100.0  & 100.0\\
           & H$\alpha$                              &296.0      & 298.4            & 298.3  & 298.2\\
           & [N\,{\sc ii}] 5755                     &0.9        & 0.7              & 0.8    & 0.8 \\
           & [N\,{\sc ii}] 6584                     &314.0      & 287.8            & 291.6  & 294.9\\
           & [N\,{\sc ii}] 6548                     &108.0      & 97.6             & 98.9   & 100.0\\
           & [O\,{\sc ii}] 3727                     &15.0       & 15.3             & 15.4   & 15.8 \\
           & [S\,{\sc ii}] 6716                     &13.2       & 17.6             & 17.1   & 17.5 \\
           & [S\,{\sc ii}] 6731                     &16.2       & 21.1             & 20.9   & 21.2\\
           & [S\,{\sc iii}] 9068                    &12.0       & 3.5              & 3.7    & 3.7\\
           & [S\,{\sc iii}] 9530                    &30.6       & 8.9              & 9.3    & 9.3\\
    &  &  &  &  & \\
           & $F$(H$\beta$) $\times$10$^{-14}$       &9.81       & 9.43             & 8.44   &  7.86\\
           &$T_\mathrm{e}$[N\,{\sc ii}]5755/6584    &6019       & 5866             & 5894   &  5909\\
           &$n_\mathrm{e}$[S\,{\sc ii}]6716/6731    &947        & 885              & 925    &  898 \\
\hline
{\it Spitzer}& H\,{\sc i} 12.3$\mu$m                & 5.2       & 5.2              & 5.2   & 5.2    \\
 IRS        & [Ne\,{\sc ii}] 12.81$\mu$m            & 165.9     & 139.2            & 137.9 & 139.1  \\
           & [Ne\,{\sc iii}] 15.5$\mu$m             & 3.1       & 0.0              & 0.0   & 0.0    \\
           & [S\,{\sc iii}] 18.7$\mu$m              & 251.5     & 47.8             & 46.0  & 45.9   \\
           & [S\,{\sc iii}] 33.5$\mu$m              & 263.8     & 37.3             & 35.6  & 35.9   \\
    &  &  &  &  & \\
           &$n_\mathrm{e}$[S\,{\sc iii}]18.7/33.5  & 554        & 922              & 936   & 922    \\
\hline
           &                                      &[W/m$^{2}$]  &[W/m$^{2}$] &[W/m$^{2}$] & [W/m$^{2}$]\\
{\it Herschel} & [N\,{\sc ii}] 121$\mu$m          & 13.1        & 18.5       & 15.0       & 14.7     \\
   PACS$^{(2)}$& [N\,{\sc ii}] 205$\mu$m          &1.9          & 2.6        & 2.2        & 2.1      \\
           & [N\,{\sc iii}] 57.2$\mu$m            &0.3          & 0.0        &0.0         & 0.0      \\
           & [O\,{\sc i}] 63.2$\mu$m              &1.3          & 0.0        &0.0         & 0.0      \\
           & [O\,{\sc i}] 145$\mu$m               &0.1          & 0.0        &0.0         & 0.0      \\
           & [C\,{\sc ii}] 158$\mu$m              & 2.1         & 5.3        & 4.3        & 4.3      \\
     \hline
     \hline
     \end{tabular}
     \caption{Emission lines considered in our models. Optical lines
       are relative to $\mathrm{H}\beta=100$ and infrared line
       measurements obtained from IRS spectra using\ the PAHFIT IDL
       routine \citep{Smith2007} are relative to H\,{\sc
         i}\,12.3$\mu$m. The errors from the observational
       measurements are statistical, of order
       $\sim5\%$. $^{(1)}$~\citet{Esteban1991}. $^{(2)}$~\citet{Vamvatira2016}.}
     \label{tab:linesout}
\end{center}
\end{table}

The simple model described above reproduces the gas low-ionization
emission reasonably well. However, the thermal dust emission resulting
from the ISM size distribution of silicate grains that are
homogeneously mixed with the gas at all radii leads to the SED shown
in Appendix~\ref{app:ismsed} (see Figure~\ref{fig:SED2}). There is a clear excess
of emission at shorter wavelengths, i.e., there is too much hot dust.

\begin{figure*}
\centering
\includegraphics[width=0.9\linewidth]{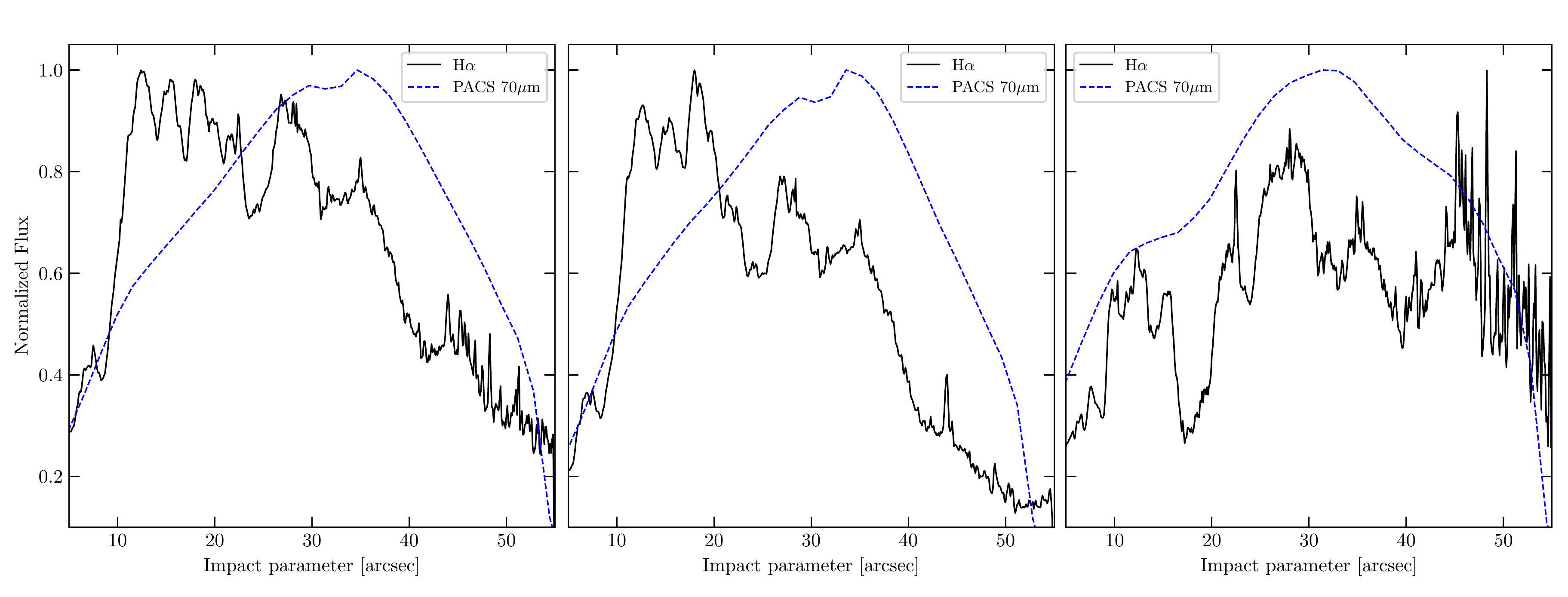}
\caption{Projected radial distribution of the surface brightness
  integrated over (a) whole nebula ($0^{\circ}\leq \phi \leq 360^{\circ}$), (b) the bipolar emission region of M\,1-67 at 70$\mu$m ($-30^{\circ}\leq \phi \leq 90^{\circ}$ and $130^{\circ}\leq \phi \leq 230^{\circ}$), and (c) regions outside of bipolar emission ($90^{\circ}\leq \phi \leq 130^{\circ}$ and $230^{\circ}\leq \phi \leq 330^{\circ}$). The solid black line is
  the observed H$\alpha$ surface brightness. The dashed blue line is
  the observed surface brightness from the \textit{Herschel} PACS band
  at 70~$\mu$m.}
\label{fig:HaHBS}
\end{figure*}

The dust temperature can be lowered by either increasing the size of
the dust grains or moving the dust further away from the
  star. To better understand the
distribution of dust in the nebula, we calculated the radial
distribution of the surface brightness of the 70\,$\mu$m emission
integrated over the bipolar shape seen in
Figure~\ref{fig:m167mw}. This has been done by using the 
{\sc cart2pol} routine\footnote{\url{https://github.com/e-champenois/cart2pol}}. The
radial distribution of the 70\,$\mu$m emission shown in
Figure~\ref{fig:HaHBS} strongly suggests that it originated in a
shell. Furthermore, the radial distribution of the H$\alpha$ surface brightness for the same bipolar region 
shows evidence for
emission from a structure at intermediate radii (see
Figure~\ref{fig:HaHBS}).

The MBB fit to the long-wavelength ($\lambda \ge 100~\mu$m)
\textit{Herschel} emission suggested that the grains responsible for
this part of the SED have a characteristic temperature of $T_\mathrm{D} =
38$~K. This grain population will represent the bulk of the mass of
grain material in the nebula and will be comprised of large grains. On
the other hand, small grains are necessary to produce the short
wavelength thermal emission, and these grains absorb a greater
proportion of the UV photons and affect the optical properties of the
nebula.

\subsection{Dust models}
\label{ssec:dust}

\begin{figure}
\centering
\includegraphics[width=0.9\linewidth]{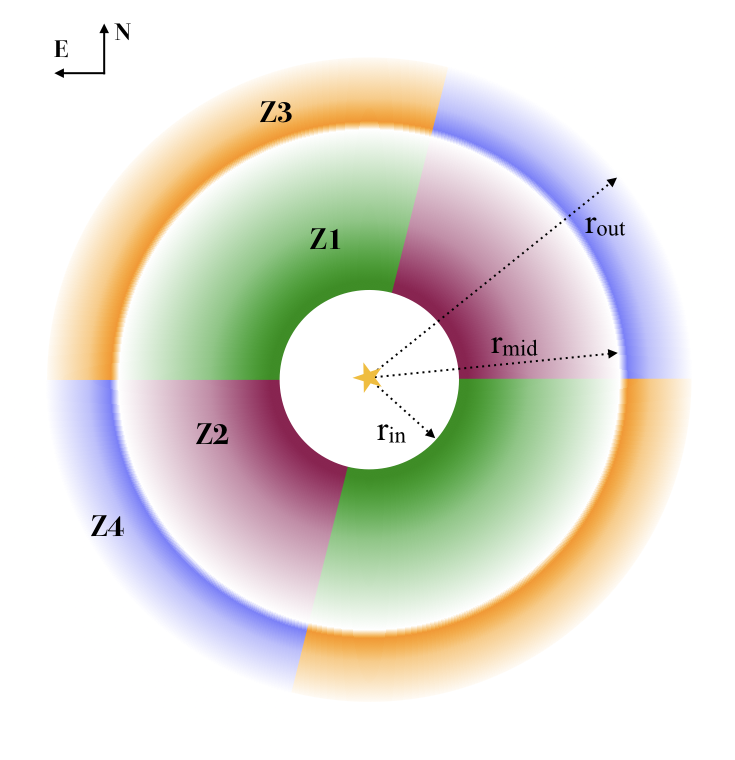}
\caption{Schematic view of the distribution of gas and dust in
  M\,1-67. Regions Z1 (green) and Z2 (purple) are composed purely of gas 
  but with differing density distributions; region Z4 (blue) is also 
  dust free and has a steeper power law to Z2; Z3 (orange) is the only 
  region that contains dust, see Table~\ref{tab:input} for details 
  of the density distributions. The white region interior to 
  $r_\mathrm{in}$ is completely empty of both gas and dust. 
  The diagram depicts a vertical slice ($y=0$ plane) of the 3D model, 
  which must be rotated through $\phi=\pi$ and oriented with 
  respect to the observer in order to produce the emission maps.}
\label{fig:esquema}
\end{figure}

\subsubsection{Two-shells model}
\label{sssec:tsm} 
  
Our second scenario proposes that M\,1-67 is composed of an inner
shell with no contribution from dust and an outer shell with gas and
two populations of grains~\citep[see e.g.][]{Gomez2018}. 
The inner dust-free shell is
delimited by $r_\mathrm{in}$ and $r_\mathrm{mid}$, whilst the outer
shell material is distributed between $r_\mathrm{mid}$ and
$r_\mathrm{out}$. The region interior to $r_\mathrm{in}$ is empty. 

We begin by setting $r_\mathrm{mid}$ to 45~arcsec, which corresponds
to the radius of the ring of expanding clumps reported by
\citet{Solf1982} and \citet{Sirianni1998}, and fix the outer radius at
$r_{\mathrm{out}} = 55$~arcsec, as before. The other parameters, that
is, $r_{\mathrm{in}}$, the power-law index in both shells
($\alpha_\mathrm{in}$,$\alpha_\mathrm{out}$), the density
normalizations ($n_{0,r_\mathrm{in}}$, $n_{0,r_\mathrm{mid}}$) 
and the filling factors ($f_\mathrm{in}$, $f_\mathrm{out}$) were
arrived at during an exhaustive process of testing different dust
grain size distributions and dust-to-gas ratios in the outer shell
while adjusting the inner shell parameters to maintain compliance with
the observational constraints listed in \S~\ref{ssec:constr}. 
We found that the total H$\alpha$ flux is most sensitive to the outer 
shell parameters, since that is where the bulk of the nebular mass 
resides, while the H$\beta$ surface brightness is most affected by 
the density distribution near $r_\mathrm{in}$ where the INT aperture 
is located.

Our best spherically symmetric two-shell model, labeled MB\,6-14, 
comprises two concentric shells: the inner, dust-free shell
is delimited by $r_\mathrm{in}=10$~arcsec and $r_\mathrm{mid}=45$~arcsec, has power-law index $\alpha=2$, density normalization $n_\mathrm{0,r_{in}}=2500$~cm$^{-3}$ and filling factor $f_\mathrm{in}=0.05$, and the outer, dusty shell is bounded by $r_\mathrm{mid}=45$~arcsec, $r_\mathrm{out}=55$~arcsec, has power-law index $\alpha=2$, density normalization $n_\mathrm{0,r_{mid}}=600$~cm$^{-3}$ and a smaller filling factor, $f_\mathrm{out}=0.037$.  The dust properties that give a good fit to
the observed IR SED combine a population of large grains of representative size
$0.9 \mu$m (`big grains') and a population with a MRN power-law size
distribution between $0.005 \mu$m and $0.05 \mu$m (`small
grains'), with a much higher proportion of big grains than small grains. It was also necessary to increase the dust-to-gas ratio above the default value in the outer shell. Details of this model are listed in the third and fourth columns of 
Table~\ref{tab:input}. 

Model MB\,6-14 does a good job of fitting the 
nebular optical and IR emission lines as well as reproducing $T_\mathrm{e}$ 
and $n_\mathrm{e}$ (see Table~\ref{tab:linesfit} and 
\ref{tab:linesout}). The total ionized
gas mass resulting from this model 
is 11.26~M$_{\odot}$ and the dust mass is 
$M_\mathrm{D}=0.22$~M$_{\odot}$.

We found that dust with sizes between 0.8--1.0~$\mu$m 
(represented, for simplicity, by a population of single-size grains of size
$0.9 \mu$m), and located in
the outer shell, has a very similar temperature to the 
characteristic dust temperature suggested by the
MBB fit, that is $T_\mathrm{D} \sim 38$~K. 
The shorter wavelength part
of the IR SED requires a population of smaller grains
with sizes in the broad range $0.005 \mu$m to $0.05 \mu$m. 
The dust in the smallest size bin is twice as hot.

\subsubsection{Bipolar shells model}
\label{sssec:bipolar}

Motivated by the evident bipolar morphology revealed 
by the IR images (see Fig.~\ref{fig:m167mw}), we introduced a further refinement to our models. \textsc{pyCloudy} enables us to use fully 3-dimensional density structures and so we relaxed our assumption of spherical symmetry and defined the bipolar density distribution depicted in Figure~\ref{fig:esquema}. The form of the density law in the different regions of the bipolar structure was guided by the surface brightness profiles depicted in panels (b) and (c) of Figure~\ref{fig:HaHBS}.

The final model, labeled MC\,6-14, 
consistent with both the optical constraints detailed
in \S~\ref{ssec:constr} and the observed IR SED, is similar to the previous model MB\,6-14 in that it requires an inner,
dust-free shell and an outer shell where the dust resides. However, in addition, both the inner and outer shells are divided into sectors having different density distributions and, moreover, part of the outer shell is dust free, to simulate the bipolar nature of the dust emission seen in the IR images.

In Figure~\ref{fig:esquema} regions Z1 (green) and Z2 (purple) are dust-free inner regions, region Z4  (blue) is a dust-free outer region, and region Z3 (orange) is the only region that contains dust. The density distribution in region Z1 is a steep power law, while that in region Z2 is much flatter, as suggested by the surface brightness profiles in Figure~\ref{fig:HaHBS}. The outer regions, Z3 and Z4, both require a steep fall-off in density but the filling factor in the dust-free sector is much lower than that in the dusty region. Full details of the parameters for this model are given in Table~\ref{tab:input}. In addition, we must specify the viewing angle and the placement of the apertures for the INT and PACS simulated emission lines. In particular, the INT aperture crosses regions Z1 and Z2, illustrated in Figure~\ref{fig:pycl}.

\begin{figure*}
   \centering
\includegraphics[width=0.9\linewidth]{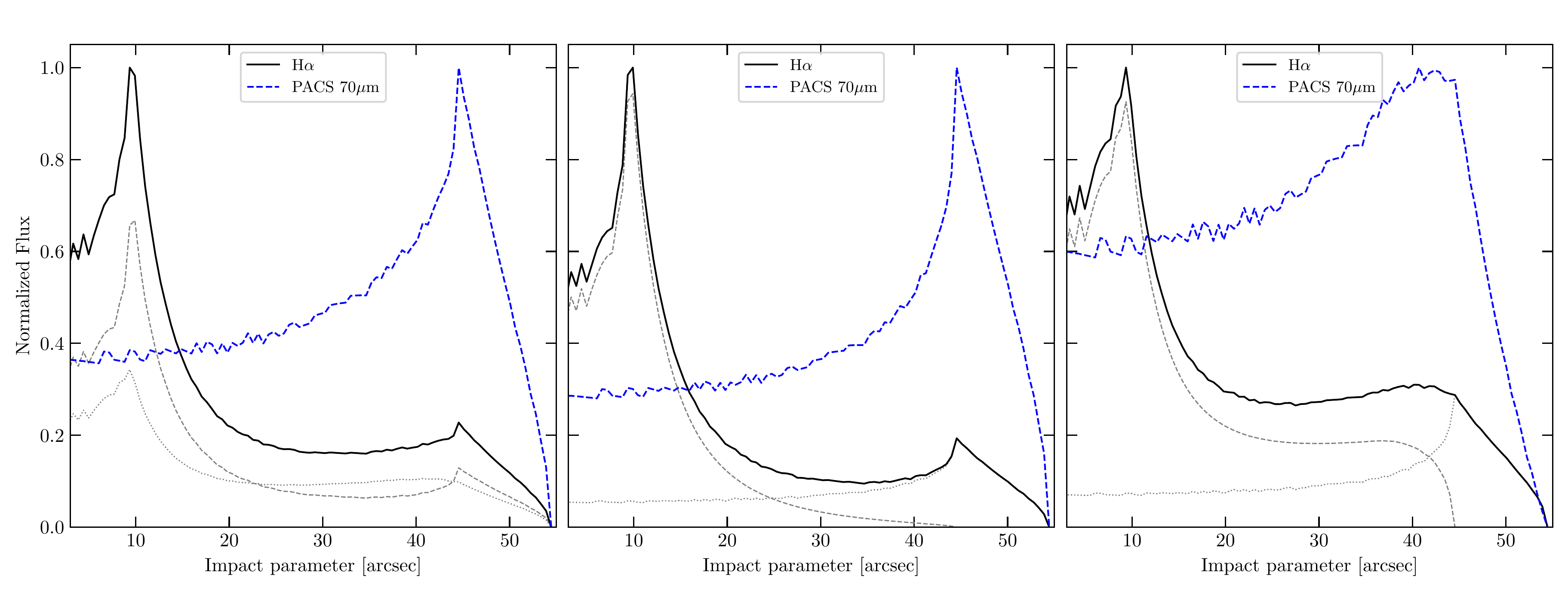}
\caption{Projected radial distribution of the surface brightness from our best model (MC\,6-14)
  integrated over (a) whole nebula ($0^{\circ}\leq \phi \leq 360^{\circ}$), (b) the bipolar emission region of M\,1-67 at 70$\mu$m ($-30^{\circ}\leq \phi \leq 90^{\circ}$ and $130^{\circ}\leq \phi \leq 230^{\circ}$), and (c) regions outside of bipolar emission ($90^{\circ}\leq \phi \leq 130^{\circ}$ and $230^{\circ}\leq \phi \leq 330^{\circ}$). The solid black line is
  the observed H$\alpha$ surface brightness. Right panel: The
dashed gray line and the dotted gray line show Zone~2+Zone~4 and Zone~1+Zone~3
contributions, respectively. Central and left panels: The
dashed gray line and the dotted gray line show inner and outer zones contributions, respectively. The dashed blue line is
  the observed surface brightness from the \textit{Herschel} PACS band
  at 70~$\mu$m.} 
\label{fig:sbprof}
\end{figure*}

\begin{figure}
   \centering
\includegraphics[width=\linewidth]{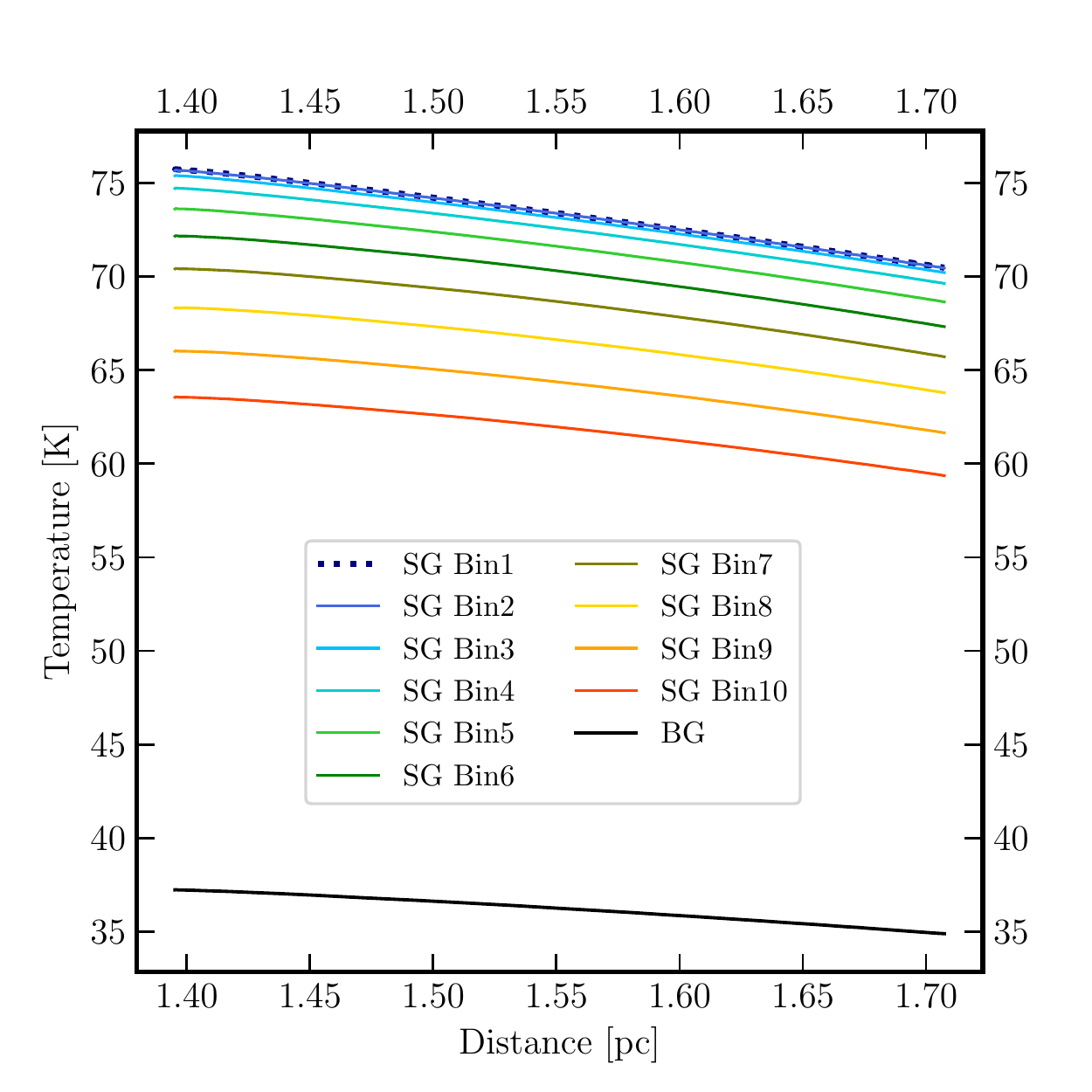}
\caption{Dust temperature profiles from our best 
model MC\,6-14. The model has 10 size bins from the population of small 
grains (SG) and only one of the big grains population (BG).}
\label{fig:dust_temp}
\end{figure}

The simulated radial H$\alpha$
surface brightness distribution of this model is shown in
Figure~\ref{fig:sbprof}. The dust properties that give a good fit to
the observed IR SED combine a population of large grains of representative size
$0.9 \mu$m (`big grains') and a population with a MRN power-law size
distribution between $0.005 \mu$m and $0.05 \mu$m (`small
grains'). As expected, these are the same populations as we found 
for the two shells model discussed above and we find the same 
proportion of big to small grains. However, now that all the 
dust is confined to a smaller volume of the model, we require 
an even higher dust-to-gas ratio to explain the observations 
(see Table~\ref{tab:input}). The resultant temperature profiles 
for the different grains in our model are shown in 
Figure~\ref{fig:dust_temp}. Variations in temperature across the 
shell are small for a given dust size bin but there are noticeable 
temperature differences between small grains and large grains.

The synthetic 70~$\mu$m radial surface brightness distribution
corresponding to these grain distributions is shown in
Figure~\ref{fig:sbprof}.  Details of the optical and infrared emission
lines of this model are presented in Table~\ref{tab:linesfit} labeled as
MC\,6-14. The synthetic SED obtained from MB\,6-14 is shown in
Figure~\ref{fig:SED} where it can be seen to compare very well with
the observed SED. In this figure we also show separately the 
contribution from the ionized gas and that from the dust. 
It can be appreciated
that the major contribution of the gas is around 12~$\mu$m and at
160~$\mu$m. The latter due to the presence of the [N\,{\sc ii}] 121~$\mu$m 
emission line
(see Table~\ref{tab:linesout}). The free-free emission from 
the gas also influences the total emission at the longest 
wavelength point at 500~$\mu$m.

We note that in order to fit the IR SED we had to increase the
dust-to-gas ratio in the outer shell ($D/G$) compared with that in the ISM
($D/G_\mathrm{ISM}$). Our best model,  MC\,6-14, 
requires a  dust-to-gas ratio 
16 times that of the ISM and a mass ratio
of big to small grains of $B/S = 20$. 
With these values, our best, self-consistent photoionization model finds 
a  dust mass of 0.22~M$_{\odot}$, with 95\% corresponding to the 
big grains, and an  ionized gas mass for M\,1-67 of 
$9.2^{+1.6}_{-1.5}$~M$_\odot$. Here, the errors in the estimated 
ionized mass were computed taking into account the errors in the 
H$\alpha$ flux reported by \citet{Grosdidier1998}.

\section{DISCUSSION}
\label{sec:disc}

Our best \textsc{pyCloudy} model successfully reproduces the IR
photometry corresponding to the thermal emission from the dust in the
nebula and approximates the main characteristics of the photoionized
gas in M\,1-67. In this section we discuss the limits of our model in
accounting for all of the observed spectral features, and also place
our model in the context of the evolution of the central star.

\subsection{Gas emission lines}
\label{ssec:elines}

The gas in the nebula absorbs the majority of the ionizing photons
emitted by the star, which correspond to wavelengths around the peak
of the SED. The shape of the incident stellar spectrum is important:
even though WNL stars have high effective temperatures,
$T_\mathrm{eff} \sim 45$~kK, the EUV photons are reprocessed to longer
wavelengths in the process of driving the strong stellar wind. The
stellar spectrum that emerges is lacking in the highest energy
photons, $h\nu > 25$~eV (see Figure~\ref{fig:pwbb} and
Table~\ref{tab:powr}). Thus, the electron temperature in the nebular
gas is low, $T_\mathrm{e} \sim 5909$~K and high ionization lines, such
as [O\,{\sc iii}]\,$\lambda 5007$, are absent.

In M\,1-67, the observed low-ionization emission features of hydrogen,
nitrogen and sulphur in the optical spectra are consistent with the
photoionization of an essentially power-law density distribution (see,
e.g., Eq.~\ref{eq:plaw}) by the WNL stellar spectrum proposed by
\citet{Hamann2006}, with the chemical abundances derived by
\citet{Esteban1991}. Our best model returns derived electron
temperatures and densities from simulated emission-line ratios (see
Table~\ref{tab:linesout}) 
within the ranges reported in the literature
\citep[e.g.,][]{Esteban1991,Fernandez2013,Sirianni1998}. However, our models are
not able to reproduce the intensities of high-ionization species, such
as [S\,{\sc iii}], which are seen in optical and \textit{Spitzer} IRS
spectra. Furthermore, the emission of the neutral [O\,{\sc i}]
63$\mu$m line detected in selected \textit{Herschel} PACS spatial
pixels \citep{Vamvatira2016} is not replicated by our models.

Although our best model does produce some [S\,{\sc iii}] emission, the
simulated intensities of the optical $\lambda\lambda$\,9068\,\AA\ and
9530\,\AA\ lines and the IR 18.7\,$\mu$m and 33.5\,$\mu$m lines are
much lower than what is observed in M1-67. In addition, our
\textsc{Cloudy} models do not produce any [Ne\,{\sc iii}] or [N\,{\sc
    iii}] emission, which is seen in the \textit{Spitzer} IRS and
\textit{Herschel} PACS spectra, respectively. In our photoionization
model, the [S\,{\sc iii}] emission is produced primarily at the inner
radius $r_\mathrm{in}$ by the most energetic photons of the incident
stellar spectrum. It is not possible to model these high-ionization
species with the adopted stellar SED (WNL 6-14), which~\citet{Hamann2006} found
to be a good fit to the optical and UV spectra of WR\,124. An SED with a greater proportion of photons able to photoionize S to S$^{++}$ may be required. Alternatively, we
speculate that the high-ionization emission lines could be produced in
shocks but such physics is not included in {\sc Cloudy}. For example, shocks could form in the 88\,km\,s$^{-1}$ bipolar flow detected by
\citet{Sirianni1998} and \citet{Fernandez2013}, which is oriented in a
northwest-southeast direction. Alternatively, the [S\,{\sc iii}] emission could arise close to
the star where the fast stellar wind shocks against the innermost
nebular material.

The other emission lines that our photoionization model does not
reproduce are those of neutral oxygen. Although \citet{Esteban1991}
only find upper limits for the optical
[O\,{\sc i}]\,$\lambda$\,6300\,\AA\ line, in the infrared the
[O\,{\sc i}]\,63\,$\mu$m is detected in several of the
\textit{Herschel} PACS spaxels, and is strongest for those that cover
bright regions of emission at 70\,$\mu$m. The 63\,$\mu$m line is
usually emitted by neutral gas in the PDR
just outside an ionization front. We speculate that this emission
comes from small, dense, unresolved neutral clumps in the nebula,
which are photoionized on their surface closest to the central
star. The [O\,{\sc i}]\,$\lambda$\,6300\,\AA\ optical line emission
that we would expect to see from the ionization front is simply too
weak to detect from such small regions. In addition, we have inspected
both the H$\alpha$ image of \citet{Grosdidier1998}, and the
emission-line maps presented by \citet{Fernandez2013} and we do not
see the characteristic photoevaporated flows coming off the starward
side of the condensations in the nebula that we might expect if the
neutral material is the main component of the condensations we see
coincident with the PACS spaxels superimposed on the 70\,$\mu$m
image. \citet{Fernandez2013} do find that the reddening coefficient
$c(\mbox{H}\beta)$ is not homogeneous across the nebula, and varies
between 1.3 and 2.5. There is no particularly extreme value associated
with any of the blobs in the 70\,$\mu$m image.

\subsection{Dust properties}
\label{ssec:dprops}

The dust size distribution predicted by our models consists of two
populations: the first has a MRN power-law size distribution with
$a_\mathrm{min} = 0.005~\mu$m and $a_\mathrm{max} = 0.05~\mu$m and a
power-law index $q = -3.5$.  The second population consists of
large grains with representative size $a_{\mathrm{big}}=0.9~\mu$m, which are necessary to
reproduce the $\lambda \geqslant 100~\mu$m \textit{Herschel}
photometry. We find that a mass proportion between big grains and
small grains of $B/S=20$ is required. The maximum size of the grains is
considerably smaller than that predicted by \citet{Vamvatira2016},
whose pure dust model required a population of large dust grains up to
10~$\mu$m in size to model the \textit{Herschel} photometric
observations. In our models, the gas absorbs and processes most of the ionizing
photons, so it is not necessary to have such large dust grains to
reproduce the dust temperature indicated by the photometry (see
\S~\ref{ssec:dust}). A maximum grain size
$a_\mathrm{big}=0.9 \mu$m is much easier to explain than
$10~\mu$m. \citet{Kochanek2011} showed that dust formation in the
winds of evolved massive stars $M_\mathrm{ZAMS} \ge
40~\mathrm{M}_\odot$ most likely occurs through eruptive events in an
LBV phase, such as the ``great eruption'' of $\eta$~Carina. This is
because the mass-loss rate $\dot{M}$ needs to be high enough to shield
the dust-formation region from soft UV photons. Mass-loss rates
$\dot{M} \ge 10^{-2}~ \mathrm{M}_\odot$\,yr$^{-1}$ are required to
obtain $a_\mathrm{big} > 1~ \mu$m. The maximum grain size of our
best-fit model is $a_\mathrm{1} = 0.9~\mu$m, which requires mass-loss
rates above $\dot{M} > 10^{-3}~ \mathrm{M}_\odot$~yr$^{-1}$ \citep[see
  fig.\,2 of][]{Kochanek2011}.  An eruptive scenario would be
consistent with the kinematic evidence of M\,1-67 presented by
\citet{Sirianni1998}, which consists of a hollow shell of clumps
expanding at 46~km~s$^{-1}$ centred on WR\,124. The second N-rich
bipolar kinematic component with expansion velocity $\sim
88$~km~s$^{-1}$ \citep{Sirianni1998,StLouis2017} is not spatially
coincident with the dust emission shown in our
Figure~\ref{fig:m167mw}.

Our estimate for the total mass of dust in M\,1-67 is $M_\mathrm{D}=0.22$~M$_{\odot}$, 
of which 0.206~M$_{\odot}$ corresponds to the largest
size grains, $a_\mathrm{big} = 0.9 \mu$m. This is $\sim$2/3 smaller 
than the dust mass
obtained from the modified black body model (see
\S ~\ref{ssec:dust}), and suggests that the standard Milky Way
normalization $\kappa_{\nu0}$ value is not appropriate for M\,1-67. Indeed,
the normalization value $\kappa_{\nu 0}$ includes the standard Milky Way
hydrogen-to-dust ratio, set to $H/D = 90$ \citep{Draine2003,Bianchi2019}, 
but we have found that a much higher proportion of dust is required in this nebula.

Increasing the mass of silicate grains by 
a factor of sixteen
as required by our model is puzzling and needs some further 
discussion, since the underlying assumption was that virtually all of the 
available silicon atoms were locked up in dust
grains \citep{Weingartner2001}. Thus, to increase the mass fraction 
we would require more
silicon atoms than are available. A possible explanation is a
nucleosynthesis event such as that proposed by
\citet{Podsiadlowski2010} through an explosive common envelope
ejection (ECEE) scenario but we calculate that the energy liberated in
fusion reactions to produce the atoms needed for the extra silicate
dust grains is $9.7\times10^{-7}$~erg~nucleon$^{-1}$. That is, the
energy to convert 0.22~$M_{\odot}$ into dust is
$2.50\times10^{50}$~erg. This amount of energy would result in the
surrounding envelope being expelled at velocities far higher than the
46\,km~s$^{-1}$ expansion velocity observed in the shell. 
This scenario is therefore unlikely.

A more viable explanation is that the silicate grains have
different optical properties to the spherical grains assumed in our
models. For example, \citet{Siebenmorgen2014} find that spheroidal
grains have far-IR absorption cross-sections a factor 1.5--3 larger
than spherical grains with the same volume. Dust masses estimated
without taking this into account would be overestimated by the same
factor. Additionally, \citet{Gomez2018} found that the emission of
spheroidal grains is shifted to longer wavelengths in comparison with
spherical grains. If the grains are fluffy, that is, porous aggregates
of smaller grains, the absorption cross-sections will be even
higher and a much smaller mass of grain material would be required 
to reproduce the observd IR SED. We intend to explore different 
grain properties in future
work.

\subsection{Mass loss and stellar evolution}
\label{ssec:new}

Our \textsc{pyCloudy} model MC\,6-14 finds the ionized gas mass in
the nebula to be $9.2^{+1.6}_{-1.5}~\mathrm{M}_\odot$. 
The estimated mass for the
central star WR\,124 is $22~\mathrm{M}_\odot$ \citep{Hamann2019},
which means that the initial mass for this star was at least
$31~\mathrm{M}_\odot$. If we take into account mass loss during 
main sequence evolution
where single stars
with ZAMS masses above 30~M$_\odot$ can lose $\sim$10~M$_\odot$
\citep[see][]{Ekstrom2012,Georgy2012}, the initial mass for WR\,124 
could easily be $> 40 M_\odot$.

We can estimate the average mass-loss rate of the star during the
dust-formation stage. Assuming a constant expansion velocity of
46~km~s$^{-1}$, the elapsed time between the ejection of the outer
($r_\mathrm{out} = 55$~arcsec) and inner ($r_\mathrm{in} = 45$~arcsec)
edges of the shell is $6.6\times10^3$~yrs, where we have taken the
distance to the nebula to be 6.4~kpc. According to our model, 
the mass of gas corresponding to 
the shell is $3.98~ \mathrm{M}_\odot$, then the mass-loss rate when the
shell was expelled is of order $\dot{M}_\mathrm{shell} = 6\times
10^{-4}~\mathrm{M}_\odot \mathrm{yr}^{-1}$. This is 
within the range of  values listed for mass-loss rates of known 
Galactic LBVs with dusty
shells \citep[see Table~1 in][]{Kochanek2011}.

We have discussed WR\,124 and its surrounding nebula in the context of
single massive star evolution. \textit{Hipparcos} light curves over
the three year period 1990 to 1993 show no evidence of periodic
variability so any binary companion would have to be very low
mass. However, \citet{Toala2018} recently speculated that the marginal
detection of hard X-rays from WR\,124 does not rule out the presence
of a compact object such as a neutron star embedded within the dense
stellar wind of the WN8 star. That means that the evolutionary 
scenario for WR\,124
would include substantial mass transfer at some point
\citep{DeDonder1997}.

In a close binary system, common envelope evolution can lead to the
ejection of the common envelope (CE), and produce a tighter 
binary \citep[][and references therein]{Ivanova2013}. Once the CE forms,
frictional drag causes the secondary component of the binary to
spiral-in and orbital energy is transferred to the envelope. If enough
energy is transferred, a substantial fraction of the envelope can be
ejected. If the gas in the envelope can recombine and form molecules
during the ejection event, the recombination energy can be an
additional driving mechanism for the envelope ejection
\citep{Ivanova2013}. If we identify the shell between 45~arcsec and
55~arcsec in our model with the observed clumpy ring expanding at
46\,km~s$^{-1}$ \citep{Solf1982,Sirianni1998}, then the kinetic energy
of the shell is of order $8\times 10^{46}$~erg, which is similar to
that estimated for outbursts in recent optical transient events that
have been associated with mass-transfer processes in binary systems
\citep{Kashi2010}. Such accretion processes can also power the
ejection of a CE. However, we estimate the timescale for
the ejection event in M1-67 to be much longer than, for example, the
Great Eruption of the massive binary system Eta Carinae, which
occurred over a twenty year period in the mid 19th century. This
suggests that the spiral-in scenario is more likely, since this can
occur over many dynamical timescales
\citep{Ivanova2013}. \citet{Iaconi2019} have modelled the asymptotic
behavior of the common envelope expansion after the end of the
dynamic in-spiral, and find that it evolves into an ordered,
shell-like structure at late times.

Finally, the ejecta of CE systems have been shown to be an ideal
environment for dust formation, which can form more efficiently in
certain CE conditions than in the envelopes of cool stars
\citep{Lu2013}. The bipolar distribution of the thermal dust emission
in M\,1-67 at all wavelengths further supports a binary system origin. 
A CE phase
is now thought to be one of the main routes to producing Type~Ib
supernovae, where the progenitors lose their hydrogen envelopes and
become Wolf-Rayet stars before exploding some $10^4$--$10^6$~years later,
depending on the stage of evolution where the mass transfer occurs
\citep{Podsiadlowski1992}.

The analysis of the IR data in combination with a self-consistent photoionization model, 
able to reproduce both the dust and nebular properties of M\,1-67, 
lead us to conclude that
this WR nebula has not formed by the classic wind-wind interaction 
scenario nor an eruptive LBV channel characteristic of single massive star evolution. 
We suggest that M\,1-67 has formed as the result of the 
ejection of a CE of a binary system. 
If this is true, this would make M\,1-67 and its progenitor star, 
WR\,124, the first 
direct evidence of the post-CE scenario in massive stars.

\section{SUMMARY}
\label{sec:summ}

We have produced a self-consistent photoionization model that fits IR
photometry and spectra from \textit{WISE}, \textit{Spitzer} (IRS and MIPS) and
\textit{Herschel} (PACS and SPIRE) of M\,1-67, the nebula around
WR\,124. The \textsc{Cloudy} photoionization code that we used allowed
us to test different gas density distributions and dust grain size
distributions. The \textsc{pyCloudy} tool enabled us to simulate the
line emission and dust photometry through synthetic apertures
corresponding to the different reference observations.  Our principal
findings are:

\begin{enumerate}
  
  \item The stellar atmosphere model 06-14 of the PoWR WNL grid that
    \citet{Hamann2006} found to be a good fit to the stellar 
    spectrum of the central star, WR\,124, is broadly consistent with 
    the nebular optical and IR emission lines in our best model but 
    underpredicts the intensities of high-ionization lines
    (see \S~\ref{ssec:elines}).

  \item We require a two-shell structure in order to model the gas and
    dust emission. The inner region contains only gas, and is required
    to reproduce the H$\alpha$ surface brightness distribution and
    total emission. The outer shell contains both dust and gas and is
    needed to reproduce the $70 \mu$m surface brightness profile and
    the dust temperature. The outer shell could correspond to the
    clumpy ring at $45$~arcsec radius, which expands at
    46~km~s$^{-1}$.  Both inner and outer shells have a power-law
    distribution of material (see \S~\ref{sssec:tsm}).

  \item The dust grain size distribution resulting from our
    photoionization model consists of two populations of pure silicate grains. The first is a
    MRN power law with $a_\mathrm{min} = 0.005~\mu$m, $a_\mathrm{max}
    = 0.05~\mu$m and power-law index $q = -3.5$. The second is a
    population of large grains with representative size  
    $a_{\mathrm{big}} = 0.9~\mu$m, where  the
    mass proportion between the big and small grain populations is 
    $B/S$ = 20. This is in stark contrast to the results of
    \citet{Vamvatira2016} who required much larger grains in their
    dust-only radiative transfer models and highlights the role played
    by the nebular gas in the absorption of UV photons from hot stars (see \S~\ref{ssec:dprops}).
    
  \item We need a silicate grain mass fraction 16 times higher than that of 
  standard ISM dust in
    order to reach the flux of the infrared photometry. We suggest that
    non-spherical grains, whose absorption cross sections are factors
    of 2 or more higher than spherical grains of identical volume,
    could explain the discrepancy in the mass of dust and the chemical
    abundances of the grain materials (see \S~\ref{ssec:dprops}).

  \item The total dust mass in the nebula resulting from our
    photoionization model ($M_\mathrm{D} = 0.22~\mathrm{M}_\odot$)
    and the dust mass resulting from fitting a modified black body
    model to the \textit{Herschel} PACS and SPIRE photometry
    ($M_\mathrm{D} = 0.36 \mathrm{M}_\odot$) can be reconciled if the
    normalization parameter $k_{\nu 0}$ takes into account the
    specific characteristics of the material in this WR nebula.

  \item The maximum grain size $a_\mathrm{big} = 0.9~\mu$m and the   
  the average mass-loss rate in the shell ejection $\dot{M} \sim 6 \times10^{-4}~\mathrm{M}_\odot$~yr$^{-1}$ support an eruptive 
  formation of M\,1-67.
    
    \item The estimated mass of photoionized gas in the nebula is 
    9.2$^{+1.6}_{-1.5}$
      M$_{\odot}$. Since M\,1-67 is located high above the Galactic
      plane ($z \sim 370$~pc), we can assume that the nebula consists
      almost entirely of material ejected from the central star,
      WR\,124. Assuming that the current mass of WR\,124 is
      22~M$_\odot$ \citep{Hamann2019} we estimate that its initial mass has to be $> 40 M_{\odot}$.

    \item We propose that M\,1-67 has been formed as a result of a 
    common envelope ejection scenario, which easily explains 
    the energetics and dust formation in the
      outer, dusty shell. This would make M\,1-67 and its progenitor
      star (WR\,124) the first observational evidence of the post-common
      envelope evolution in massive stars.
       
\end{enumerate}

\section*{Acknowledgements}

The authors would like to thank the referee, Anthony P. Marston, 
for a critical reading of our manuscript and suggestions that 
improved our model and the presentation of the results.
The authors are grateful for financial support provided by
Direcci\'on General de Asuntos del Personal Acad\'emico, Universidad
Nacional Aut\'onoma de M\'exico, through grants Programa de Apoyo a
Proyectos de Investigaci\'on e Inovaci\'on Tecnol\'ogica 
IA100720 and a IN107019. PJH also thanks Consejo Nacional de Ciencias y
Tecnolog\'{\i}a, M\'exico for a research studentship. This work has
made extensive use of NASA's Astrophysics Data System. The authors 
thank  Christophe Morisset for helping with the implementation 
of pyCloudy and a critical reading of the manuscript.
This work makes use
of {\it Herschel}, {\it Spitzer} and {\it WISE} IR observations. 
{\it Herschel} is an ESA space observatory 
with science instruments provided by European-led Principal Investigator 
consortia and with important participation from NASA. The
{\it Spitzer} Space Telescope was operated by the 
Jet Propulsion Laboratory, California Institute of Technology 
under a contract with NASA. Support for this work was provided 
by NASA through an award issued by JPL/Caltech. {\it WISE} is a joint 
project of the University of California (Los Angeles, USA) 
and the JPL/Caltech, funded by NASA.

\section*{Data availability}

The data underlying this article will be shared on reasonable 
request to the corresponding author.



\bibliographystyle{mnras}
\bibliography{wr124} 


\appendix

\section{Distance to WR\,124}
\label{app:distance}
\begin{figure}
\centering
\includegraphics[width=0.9\linewidth]{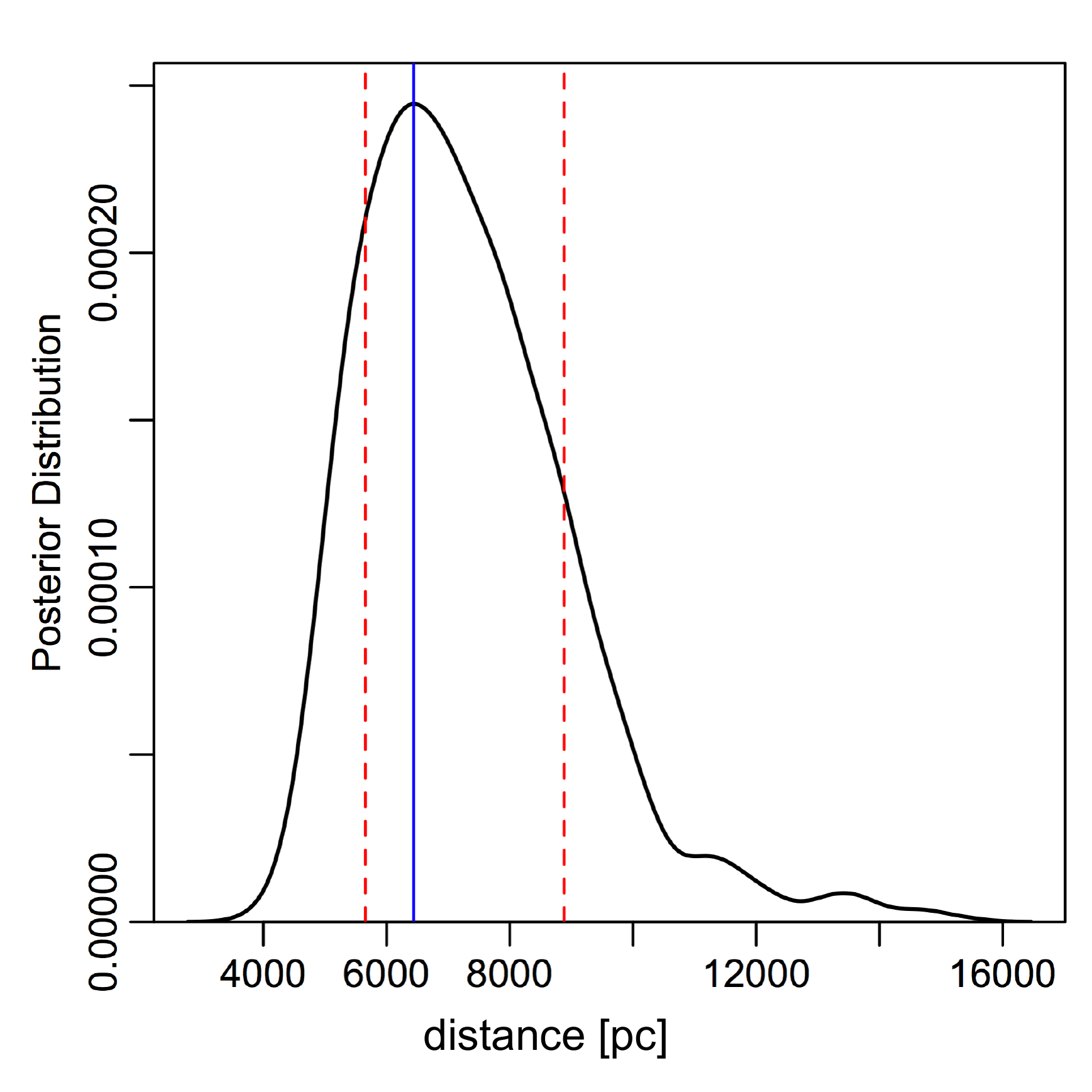}
  \caption{Posterior distribution for WR\,124. The peak of the curve
  (blue vertical line) corresponds to the most likely distance to the
  star, $6.4^{+2.45}_{-0.80}$~kpc, and the red dashed lines denote the
  one-sigma errors.}
  \label{fig:distance}
\end{figure}

Recently, \citet{Rate2020} used \textit{Gaia} Data Release 2 (DR2)
parallaxes and Bayesian methods to obtain distance estimates to 383
Galactic Wolf-Rayet stars. Their Bayesian prior relates these massive
stars to the Galactic distribution of H\,{\sc ii} 
regions and also takes into
account extinction using a Galactic dust model. \textit{G}-band
photometry is used to correct potential underestimates of the parallax
uncertainty. For WR\,124 the uncorrected parallax uncertainty is
$\sigma_\omega = 0.036$~mas, while the uncertainty taking into account the
$G$-band magnitude is not very different, $\sigma_\omega = 0.045$. Using this method, \citet{Rate2020} obtained a distance $5.9^{+1.48}_{-1.08}$~kpc to WR\,124.

For the runaway Wolf-Rayet star WR\,124, which is not associated with
any H\,{\sc ii} region and is far from the Galactic plane ($b > 3^\circ$),
choosing a prior that is based on the Galactic distribution of H\,{\sc ii}
regions does not obviously have any advantages compared to the purely
geometric model of \citet{BailerJones2018}, where the prior varies
smoothly as a function of Galactic longitude and latitude according to
a model for the distribution of stars in the Galaxy.

In this paper, we estimate the distance to WR\,124 using the
\textit{Gaia} DR2 parallaxes and also the proper motions, and employ Bayesian
methods that take into account the correlations between the parallax,
proper motions and their uncertainties \citep{Luri2018}. We correct
for the parallax zero point using the measured QSO global zero point
value of $-0.029$~mas \citep{Luri2018}. We use the Python and R-code
routines from the tutorial\footnote{\url{https://github.com/agabrown/astrometry-inference-tutorials/tree/master/3d-distance}} described in Section~5.4 of
\citet{Luri2018},	
and include the prior over distance $P(r)$ given by the geometric
model of \citet{BailerJones2018} mentioned above. The posterior
distribution is sampled numerically using an MCMC algorithm and the
results are shown in Figure~\ref{fig:distance}.

The most likely	distance to WR\,124 corresponds	to the peak of the
posterior distribution, $d = 6.4^{+2.45}_{-0.80}$~kpc, and the
one-sigma errors (indicated by the red dashed lines in
Fig.~\ref{fig:distance}) give the credible interval of distance
values. Our estimated distance is consistent, within the
one-sigma error range, with the \citet{Rate2020} value.

\section{Emission from ISM size distribution of silicate grains}
\label{app:ismsed}
\begin{figure}
   \centering
\includegraphics[width=\linewidth]{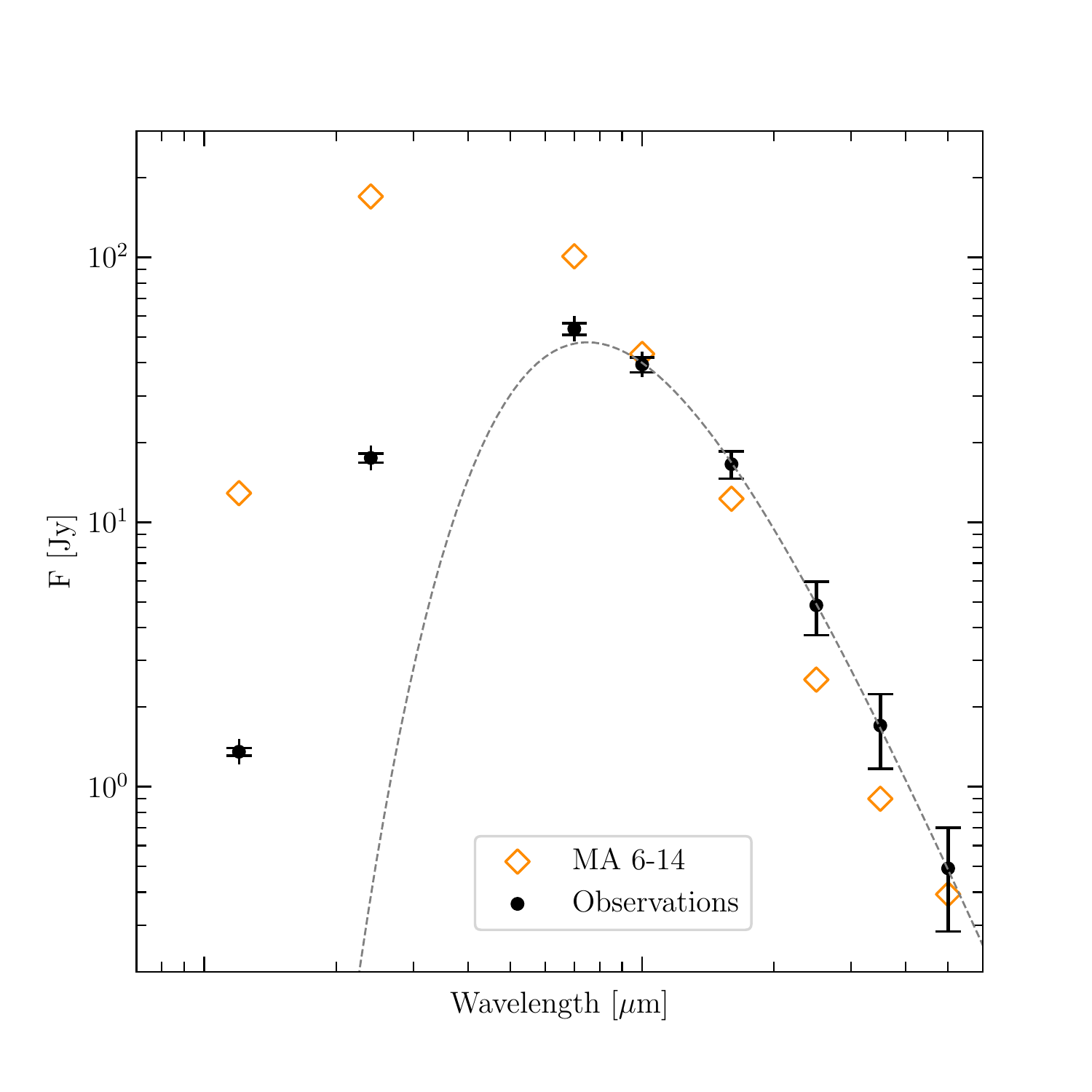}
\caption{SED obtained from the IR observations of M\,1-67 (black) dots
  (see Table~\ref{tab:fot}). The error bars take into account
  uncertainties associated with the instrument calibration and the
  background subtraction process. The synthetic SED obtained from our
  model MA\,06-14, is also shown with empty green diamonds (see
  Section~4.2 for details).}
\label{fig:SED2}
\end{figure}

Model MA 6-14 (see~\S~\ref{ssec:dust}), which has a standard ISM distribution of silicate grains, results in the SED shown in Figure~\ref{fig:SED2}. There is an excess of IR emission at short wavelengths, which is due to hot small dust grains.

\end{document}